\documentclass[journal=apchd5,manuscript=article,layout=twocolumn]{achemso}
\usepackage{graphicx}
\usepackage{cuted}
\usepackage{chemformula} 
\usepackage[T1]{fontenc} 

\author{Adrian Agreda}
\author{Sviatlana Viarbitskaya}
\affiliation[ICB UMR 6303 CNRS, Universit\'e de Bourgogne Franche-Comt\'e]
{Laboratoire Interdisciplinaire Carnot de Bourgogne, UMR 6303 CNRS, Universit\'e de Bourgogne Franche-Comt\'e, 9 Avenue Alain Savary, 21000 Dijon, France}
\author{Igor V. Smetanin}
\author{Alexander V. Uskov}
\affiliation{Lebedev Physical Institute, Leninsky pr. 53, 119991 Moscow, Russia}

\author{G\'erard Colas des Francs}
\author{Alexandre Bouhelier}
\affiliation[ICB UMR 6303 CNRS, Universit\'e de Bourgogne Franche-Comt\'e]
{Laboratoire Interdisciplinaire Carnot de Bourgogne, UMR 6303 CNRS, Universit\'e de Bourgogne Franche-Comt\'e, 9 Avenue Alain Savary, 21000 Dijon, France}
\email{alexandre.bouhelier@u-bourgogne.fr}

\title{Electrostatic control over optically-pumped hot electrons in optical gap antennas}

\let\oldmaketitle\maketitle
\let\maketitle\relax
\begin{document}
\small

 \twocolumn[
  \begin{@twocolumnfalse}
  	\oldmaketitle
    \begin{abstract}
      We investigate the influence of a static electric field on the incoherent nonlinear response of an unloaded electrically-contacted nanoscale optical gap antenna. Upon excitation by a tightly focused near-infrared femtosecond laser beam, a transient elevated temperature of the electronic distribution results in a broadband emission of nonlinear photoluminescence (N-PL).  We demonstrate a modulation of the yield at which driving photons are frequency up-converted by means of an external control of the electronic surface charge density.  We show that the electron temperature and consequently the N-PL intensity can be enhanced or reduced depending on the command polarity and the strength of the control static field. A modulation depth larger than 100\% is observed for activation voltages of a few volts. 
    \end{abstract}
  \end{@twocolumnfalse}
  ]



Plasmonic metal nanoantennas have been attracting an ever growing attention due to their ability to strongly enhance electromagnetic fields at the nanoscale and consequently improve weak nonlinear optical processes~\cite{Kauranen12}. The development of advanced simulating tools~\cite{Martin15} combined with modern nanofabrication facilities~\cite{Celebrano15} undoubtedly pushed the general understanding to a level where nonlinear plasmonics can be deployed as a device technology vehicle. From the microscopic point of view, the nonlinear responses occurring in metal optical antennas, subject to an ultrafast intense electromagnetic pulse, arise from the swift generation of an out-of-equilibrium electron distribution~\cite{Hache85,boyd86} and its associated thermalization dynamics~\cite{Voisin04,Palpant18}. Externally controlling these parameters is a difficult task. Therefore, nonlinear behaviors are essentially adjusted by engineering the strength of the local electromagnetic field either by varying the geometry during the nanofabrication stage~\cite{Berthelot12b,Viarbitskaya2013} or by optimizing the pumping conditions~\cite{Comin18,VanHulst18}. While the latter has the unique benefit to dynamically manipulate coherent nonlinear signals at ultrafast time scales, it is certainly not a viable approach for integrating nonlinear plasmonic functional units in a photonic circuit architecture. A complementary alternative strategy is to control the nonlinear activity by an electrical command applied to the system. A few demonstrators were reported in the literature where the generation of second harmonic light was tuned by plasmonic gaps loaded with nonlinear electro-optical materials~\cite{Fiorini08,Cai11}.

Our work brings an additional stone in this general context. Upon excitation by a tightly focused femtosecond near infrared pulsed laser, Au nanostructures are known to react nonlinearly by producing, among other nonlinear processes, second-harmonic light~\cite{Wokaum81,Bachelier10} and a broad up-converted photoluminescence spectrum~\cite{Beversluis03}. The latter was originally attributed to a two-photon interband absorption process~\cite{Imura04,Biagoni09} because of its near quadratic power dependence. However, alternative explanations were recently proposed including inelastic transitions~\cite{Huang14,Baumberg15}, higher-order nonlinearities~\cite{Karki18} and radiative decay of a transient out-of-equilibrium electron distribution~\cite{Haug15}. The pump laser transfers energy to the electrons in the metal producing a non-Fermi distribution for a few femtoseconds. Subsequently, electron-electron scattering leads to thermalization of the electron gas which can reach several thousands of Kelvins~\cite{Palpant18}. Before the lattice and the electron gas reach equilibrium in picoseconds time range~\cite{Nordlander15,DeAbajo16}, electrons will remain hot and possibly emit a thermal electromagnetic continuum. The presence of a hot electron distribution and its different characteristic relaxation times were clearly identified in the mechanism leading to  nonlinear photoluminescence (N-PL) emission~\cite{Dujardin15,Demichel16}. This broadband N-PL and its associated non-equilibrium electron distribution is ubiquitous of metal nanostructures and has been observed for a wide range of geometries ranging from simple resonant nanoparticles~\cite{bouhelier05PRL} to coupled antennas~\cite{hecht05sciences,Brida15} and extended two-dimensional forms~\cite{cuche14}.

We demonstrate in this report that a command voltage can enhance or reduce the nonlinear photoluminescence produced by electrically-contacted Au optical gap antennas without relying on the co-integration of nonlinear dielectrics~\cite{engheta08}; in other words, the interstice of the antennas remains unloaded. This is achieved by adjusting the electronic temperature responsible for the N-PL via an electrical control of the surface electron density. 


\begin{figure}[h!]
\centering
\includegraphics[width=3.25in]{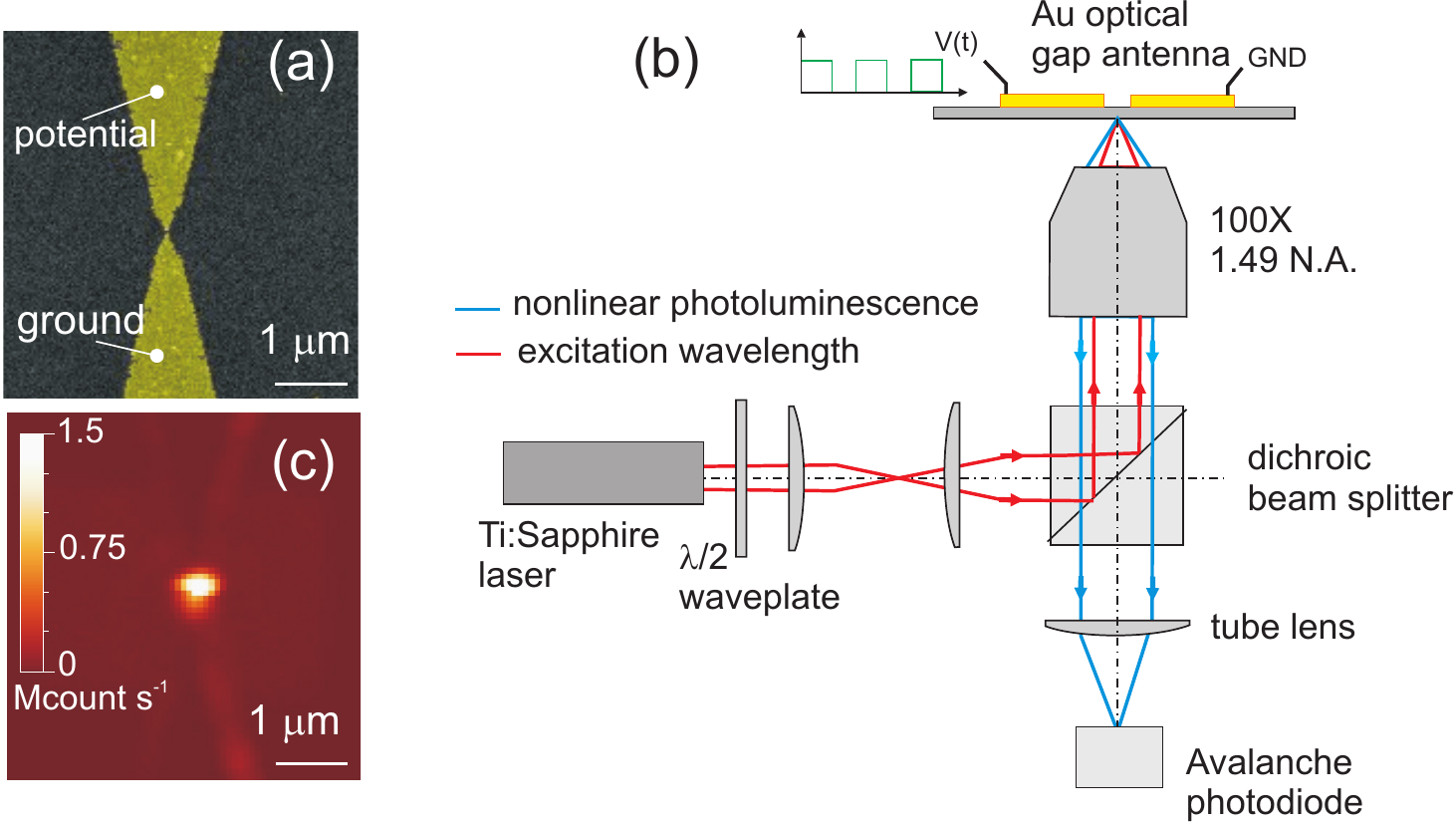}
\caption{(a) Colorized scanning electron microscope (SEM) image of a Au optical 50~nm gap antenna used in the experiment. Au is colorized by a yellow hue. (b) Experimental setup. A Ti : Sapphire laser emits 180 fs pulses at $808$ nm. The laser is focused on the optical gap antenna, which is electrically connected to a voltage source. The nonlinear photoluminescence is detected by an avalanche photodiode.(c) Confocal scanned image of the nonlinear photoluminescence emitted by the device. The area approximately covers the image in (a). }
\label{setup}
\end{figure}

Electrically-contacted gap antennas taking the form of two gold tapers separated by a sub-100~nm gap are produced by a two-step lithography.  Figure~\ref{setup}(a) shows a scanning electron micrograph of a 50 nm gap antenna produced by electron beam lithography in a first step. In the second lithography step, the microscopic electrodes contacting individual optical antennas are fabricated by photolithography. Details about the fabrication process are presented in Sect.1 of the Supporting Information. 

The sample is then transferred to an inverted optical microscope (Nikon, Eclipse) equipped with a two-dimensional piezoelectric sample scanner (Mad City Labs, NanoLP-100). A simplified sketched of the setup is pictured in Fig.\ref{setup}(b). A high numerical aperture (N.A.) objective ($\times 100$, N.A.=1.49) focuses on the sample the light of Ti:Sapph laser (Coherent, Chameleon) emitting 180~fs long pulses at a repetition rate of 80~MHz and tuned at a wavelength of 808~nm. The nonlinear up-converted photoluminescence of the sample is collected by the same objective and is spectrally discriminated from the incident wavelength by a dichroic beamsplitter and a bandpass filter. Except otherwise specified, the second-harmonic signal simultaneously generated along side the N-PL is also rejected by a notch filter centered at 405 nm.  The two leads defining the gap of the optical antenna are electrically connected to a function generator delivering a square command voltage $V(t)$. Considering the resistance of the electrical connections ($\sim 100~\Omega$), the electrical potential is mostly dropped at the interstice. For an approximately $50$~nm gap size formed by two plane-parallel plates, the electric field across the two leads is in excess of  $E=2\times10^7$~V$\cdot$m$^{-1}$ per applied volt. The photoluminescence signal locally emitted by the device is detected confocally by a photon counting avalanche photodiode module (Excelitas, SPCM-AQRH)  either as a function of the $(x,y)$ position of the sample with respect to a fixed focus spot, or as a function of the applied voltage $V(t)$ for a given position of the sample.

\begin{figure}[h!]
\centering
\includegraphics[width=3.25in]{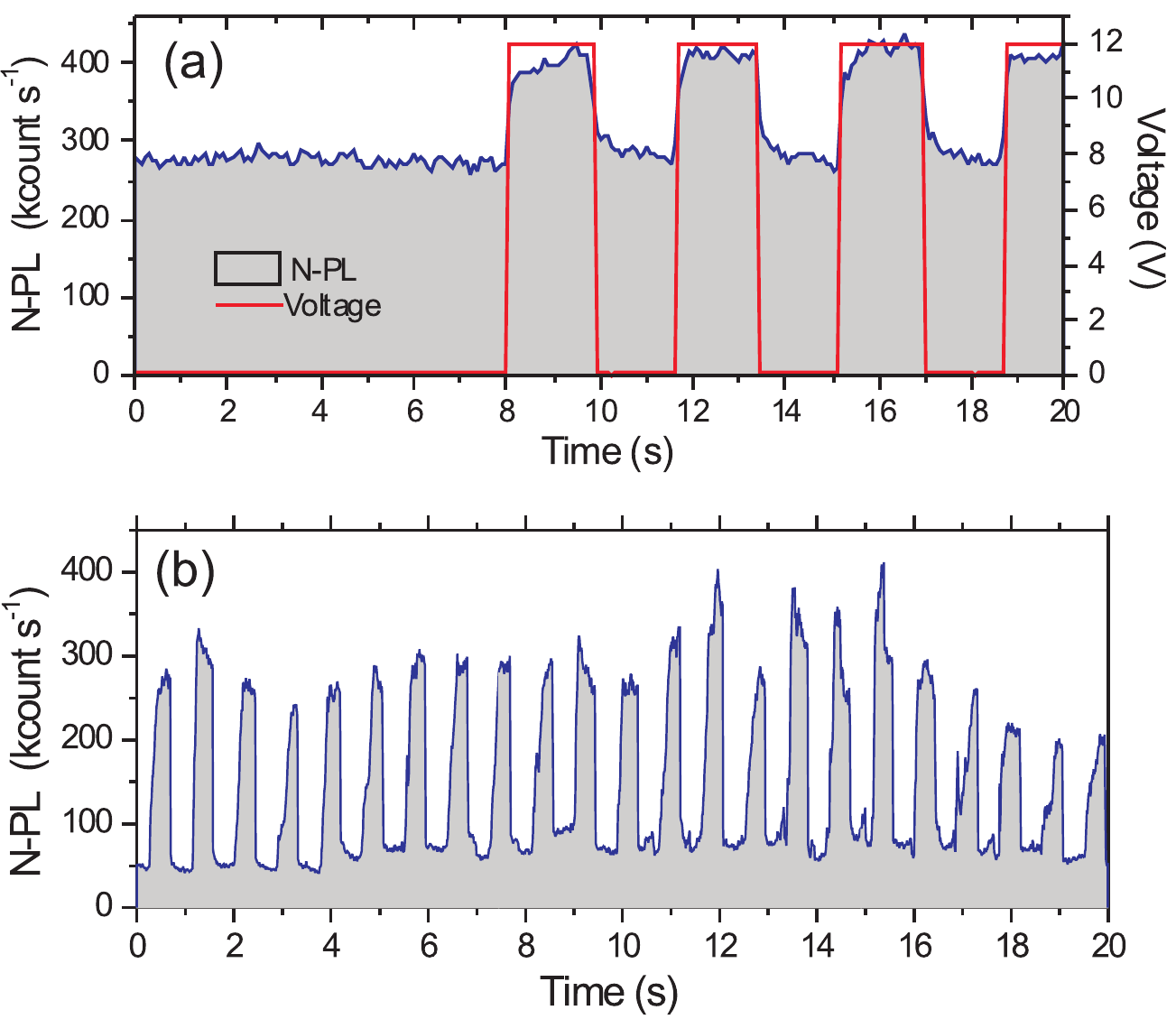}
\caption{ (a) Time traces showing the response of the N-PL to a 12~V square waveform with a period of 2~s. (b) Another N-PL time trace obtained on a different antenna with a 20~V square waveform and a period of 800~ms. The N-PL enhancement is approximately 500~$\%$. } 
\label{modulation}
\end{figure}
An example of the nonlinear $(x,y)$ photoluminescence map is illustrated in Fig.~\ref{setup}(c) for a null bias. Here the optical gap antenna is laterally scanned through the focus of the laser and the N-PL photon counts is integrated over the emitted spectrum (low pass filter at a cut-off wavelength of 680~nm). The average incident power is 290~$\mu$W and the polarization is aligned with the main axis. As anticipated, the highest N-PL signal occurs at the gap of the antenna where the incident field is enhanced~\cite{hecht05sciences}. The edges of the electrodes are also visible as a faint background in the image. As expected, the N-PL emission yield evolves nonlinearly with the incident laser power as shown in Sect. 2 of the supporting information.

Figure~\ref{modulation}(a) shows an example of N-PL voltage-induced modulation when the laser is situated closed to the antenna gap. Here, a 12~V square voltage waveform is applied to the electrical contacts at time $t=8$~s.  The voltage signal is recorded with a limited number of samples due to a limitation of the acquisition electronic when monitoring simultaneously the voltage and the APD counter. At null voltage, the N-PL count rate is constant at 270 kcount$\cdot$s$^{-1}$ indicating a residual lateral shift from the pixels featuring the highest count rate at the gap center ($\approx 520~{\rm kcount\cdot s^{-1}}$). When the voltage is applied, the N-PL rate increases to reach approximately 400~${\rm kcount \cdot s^{-1}}$, an on-state enhancement of 1.5 fold. The rate goes back to its zero-bias off-state level between two voltage pulses with an apparent delay. Further discussion will address this point later on. The enhancement factor depends on extrinsic parameters such as the applied voltage and the laser power as discussed later, but also on intrinsic figures, essentially linked to the fluctuations of the geometry of the optical gap antenna imparted by the nanofabrication process.  As an illustration of the above arguments, Fig.~\ref{modulation}(b) shows a significantly greater modulation depth obtained from a 100~nm interstice, a 20~V square waveform with a period of 800~ms and an average laser power of 49~$\mu$W. Although there is a small deviation in the N-PL maximum rate over time, the on-state enhancement reaches nearly 500~$\%$ in this experiment despite a large inte-electrode spacing.  

\begin{figure}[h!]
\centering
\includegraphics[width=3.25in]{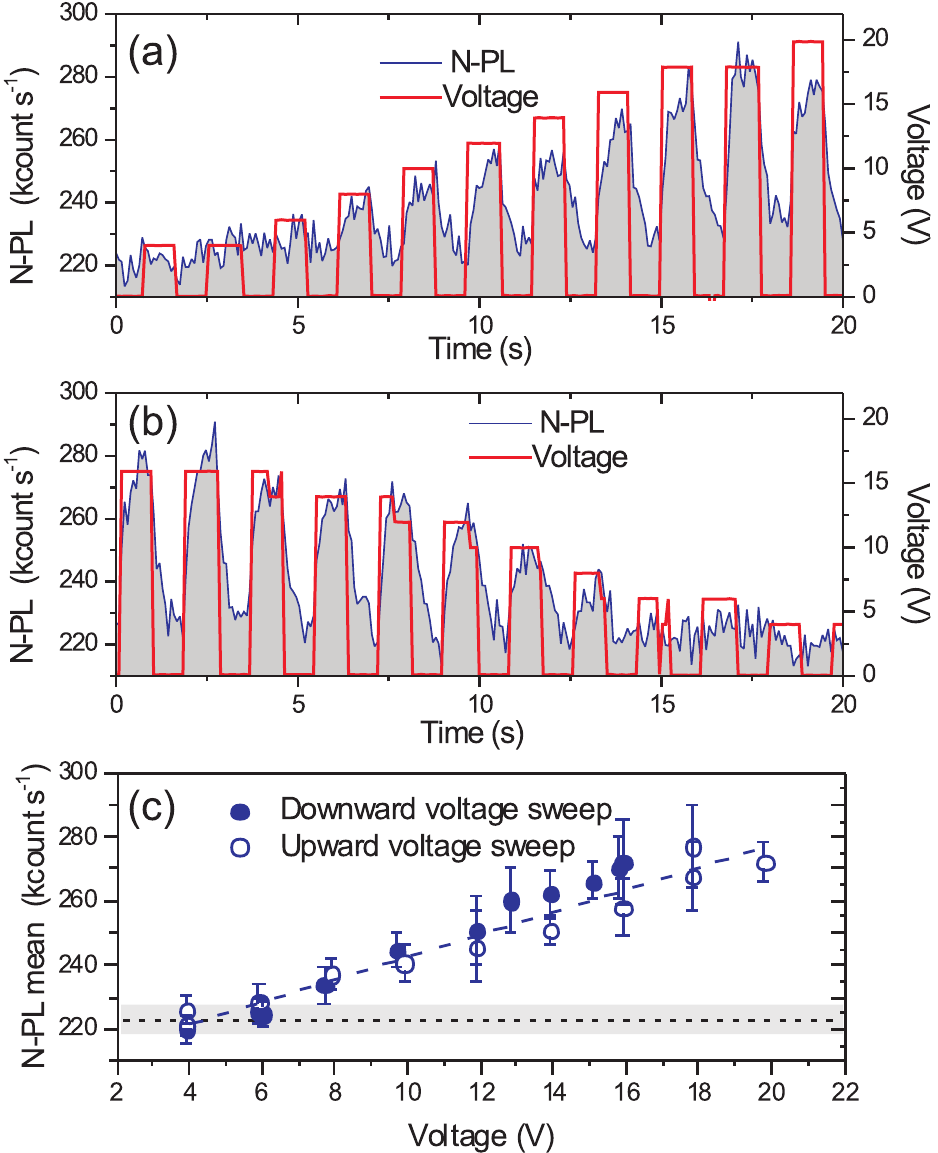}
\caption{(a) and (b) are time traces of N-PL rates responding to a change of the voltage amplitude. (c) Evolution of N-PL mean rate during the on-state as a function of the voltage applied for the two sweeps illustrated in (a) and (b). The horizontal dotted line and the grey zone indicate the mean off-state N-PL value and its standard deviation evaluated from $t<4$~s in (a) and $t>16$~s in (b)  when no modulation is detected. }
\label{voltagesweep}
\end{figure} 

In the following sections, we investigate the influence of the extrinsic control parameters on the modulation of the N-PL rate.  The parameters can be controlled with greater precision than intrinsic parameters such as a gap size or electrode morphology, which are plagued by inherent fabrication uncertainties. We start by studying the response to the amplitude of the control voltage  $V(t)$. Figures~\ref{voltagesweep}(a) and \ref{voltagesweep}(b) show time traces of the N-PL rate when the amplitude of the applied square waveform is varied in an upward and downward sweep, respectively. The data obtained during the sweeps are reproducible indicating that neither the voltage range sampled nor the laser power irreversibly degraded the optical gap antenna.  
At that stage of the discussion, one can draw preliminary conclusions. The graphs unambiguously show that the N-PL flux is sensitive to the applied voltage bias, providing thus an external electrical handle to manipulate the nonlinear activity.  Since the N-PL sits on relatively large offset at zero bias, the voltage-induced modulation becomes visible only after a certain voltage value. In the present experiment,  the modulation remains undetected  for $V<5$~V. Figure~\ref{voltagesweep}(c) illustrates the voltage dependence of the mean N-PL rate estimated during the on-state of the voltage duty cycle. A quasi linear trend is observed. The proportionality between the amplitude of the bias and the rate of N-PL is  a parameter which may prove useful for realizing a commanded device. The dotted line at 222 kcount$\cdot$s$^{-1}$ is the mean flux of the N-PL  calculated in the time sequence when the signal is unperturbed by the voltage. 

\begin{figure}[h!]
\centering
\includegraphics[width=3.25in]{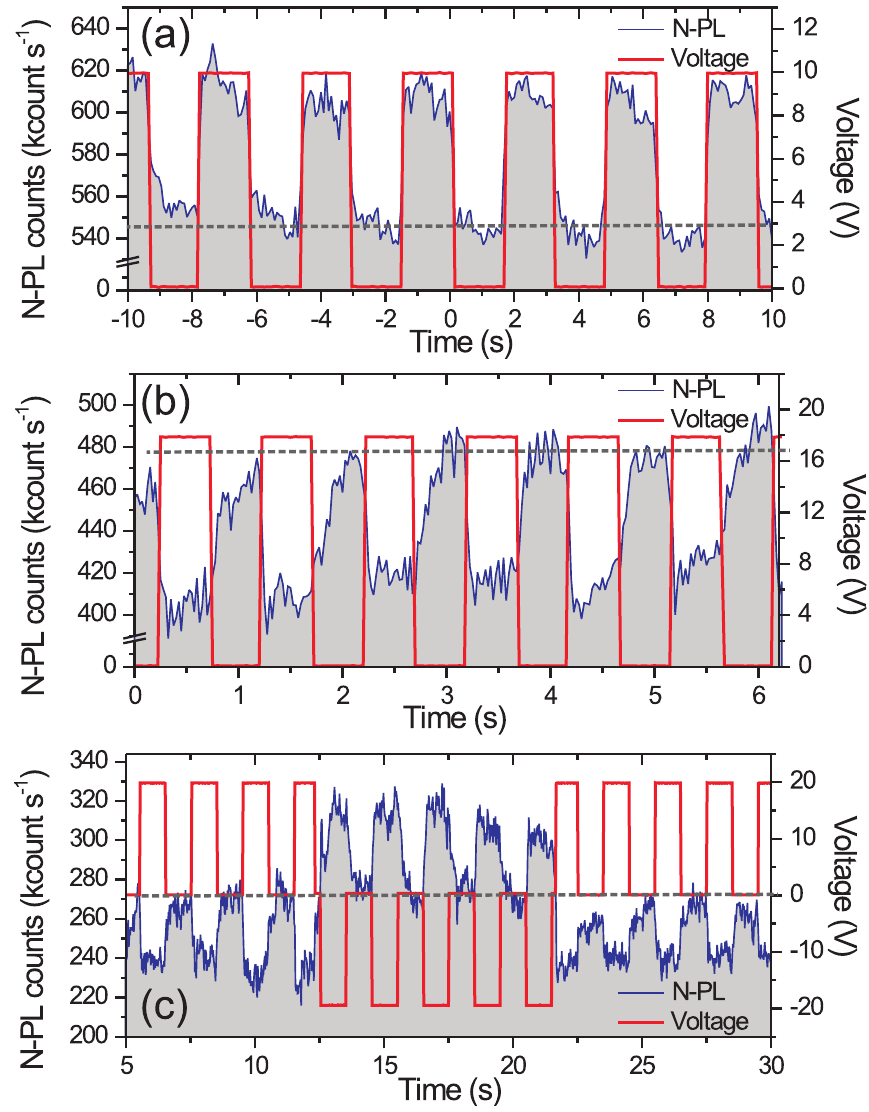}
\caption{(a) Time traces of the N-PL rate and applied bias when the laser excitation area is displaced toward the positive electrode. The N-PL and the voltage are in-phase with an enhancement of the N-PL rate on the on-state. (b) Time traces of the signals when the laser excitation is favoring N-PL emitted by the ground lead. The signals are now out-of-phase. The N-PL is partially quenched during the on-state. The dashed lines represent the approximate N-PL rate at null bias. (c) Time trace of the N-PL when the polarity of the square waveform is alternating. The N-PL undergoes out-of-phase quenching and in-phase enhancement depending on the voltage polarity.}
\label{polarity}
\end{figure}

Crucial to the observation of the N-PL modulation is the precise location of the laser spot with respect to the antenna gap. Typically, for a given bias voltage and gap size, the strength of the modulation strongly depends on the adjustment of the diffraction-limited observation area. Centering the gap in the laser focus does not necessarily render the strongest modulation. Likewise, no modulation is observed when probing the N-PL at the edges of the tapered section. We find that the modulation is generally the strongest when the tips of the tapers forming the bow-tie are  selectively illuminated. Figure~\ref{polarity}(a) and ~\ref{polarity}(b) show the time evolution of the N-PL signal when the sample is displaced from the maximum response of the gap by approximately 100 nm on either side. In this experiment, the antenna gap is nominally 80 nm. In Fig.~\ref{polarity}(a), the displacement of the sample with respect to the focus favors a N-PL signal partially overlapping the positive electrode and one retrieves the behavior observed previously where the on-state enhances the N-PL signal. In Fig.~\ref{polarity}(b), the sample is slightly moved with respect to the laser focus to favor a N-PL response from the grounded electrode. Clearly, the N-PL rate and the amplitude are now out-of-phase. The dotted lines in the graphs indicate the approximate off-state N-PL rate (null voltage). Upon application of the voltage bias, the N-PL is no longer enhanced but suffers from a significant reduction of its rate compared to its steady-state value at $V=0$~V. This behavior suggests that the polarity of the electrode contributing to the photoluminescence signal plays an important role in the control mechanism. We confirmed the above observation by changing the polarity of the amplitude without changing the position of the sample (overlap over the ground electrode). Results are shown in Fig.~\ref{polarity}(c). Again, the dashed line at 270~kcount$\cdot$s$^{-1}$ is the N-PL rate for $V=0$~V. Clearly, an on-state positive bias (here $V=+20$~V) quenches the N-PL while a on-state negative bias ($V=-20$~V) gives rise to an enhanced nonlinear signal.

\begin{figure}[h!]
\centering
\includegraphics[width=3.25in]{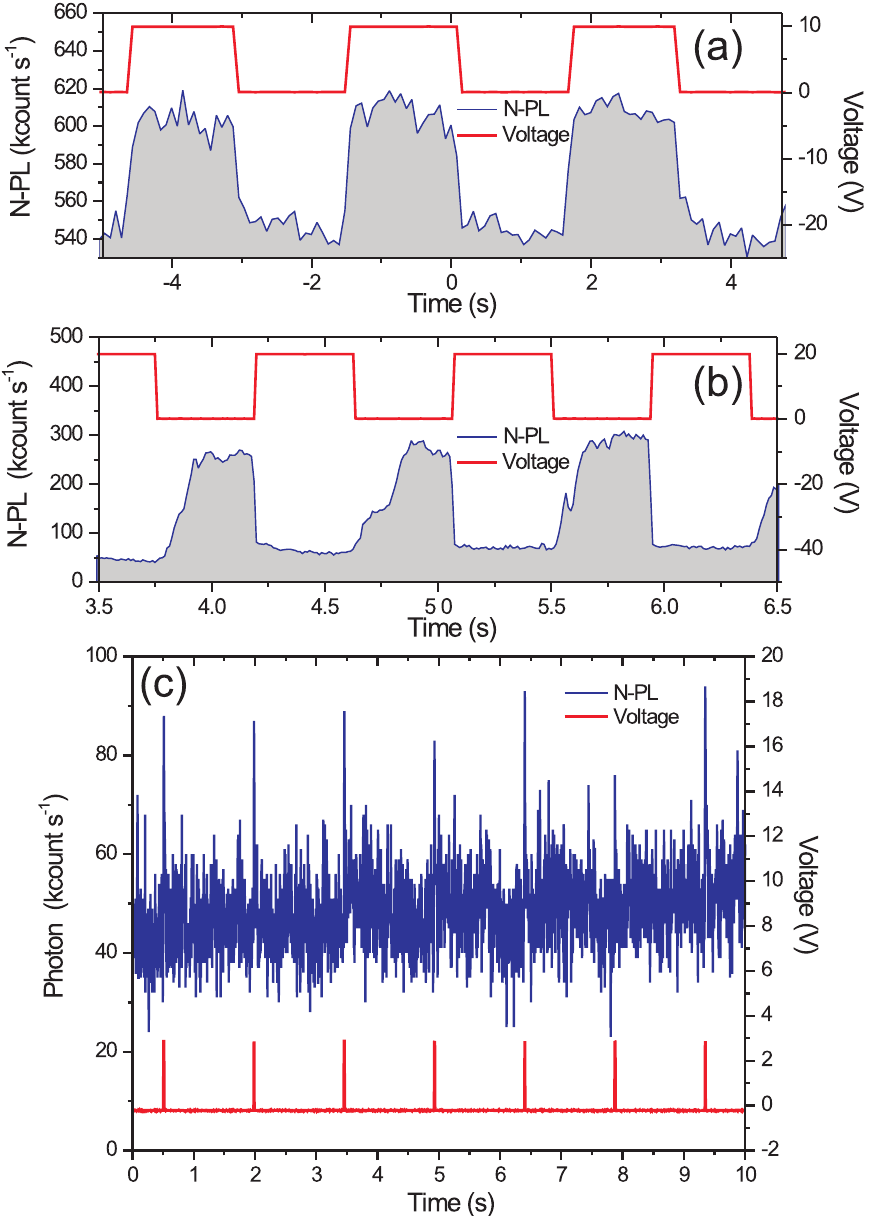}
\caption{(a) Example of rise and decay kinetics for on-state enhancement. The decay of the N-PL when $V=0$~V is 2.6 times slower than the rise time. (b) The asymmetry is reversed for on-state quenching, but the slower kinetics remains associated with the settling to null bias.  (c) Time traces of the N-PL signal and $V$ for 10 ms voltage pulses.}
\label{pulse}
\end{figure}

Enhancement or quenching of the N-PL rate when the voltage is turned on and off is not immediate. The time traces in Fig.~\ref{modulation}, Fig.~\ref{voltagesweep} and Fig.~\ref{polarity} show that the temporal evolution of the N-PL is occurring with a slow dynamics and is not symmetric with respect to the voltage edge. In Fig.~\ref{pulse}(a)   the average rise time $\tau_r$ and decay time $\tau_d$  measured at 10$\%$ and 90$\%$ of the modulation are $\tau_r=150\pm 30$~ms and $\tau_d=390\pm 80$~ms. However, the figures are here indicative. We have observed a large variability from antenna to antenna, but also between positions of the sample with respect to the laser focus. When the on-state quenches the signal, the asymmetry reverses as illustrated in Fig.~\ref{pulse}(b). However, the slower time is always associated to the settle down of the N-PL when the voltage goes to 0~V suggesting that a diffusive process is at play. If the kinetics of the N-PL signal would be mainly driven by such a slow time response, electrical signals with a duration smaller than the $\tau_r$ would not significantly affect the N-PL rate. Figure~\ref{pulse}(c) displays an experiment where the voltage bias is a 10~ms pulse. The modulation here remains substantial ($\sim 90\%$) suggesting that the kinetics is driven by a combination of a fast and a slow process.

\begin{figure}[h!]
\centering
\includegraphics[width=3.25in]{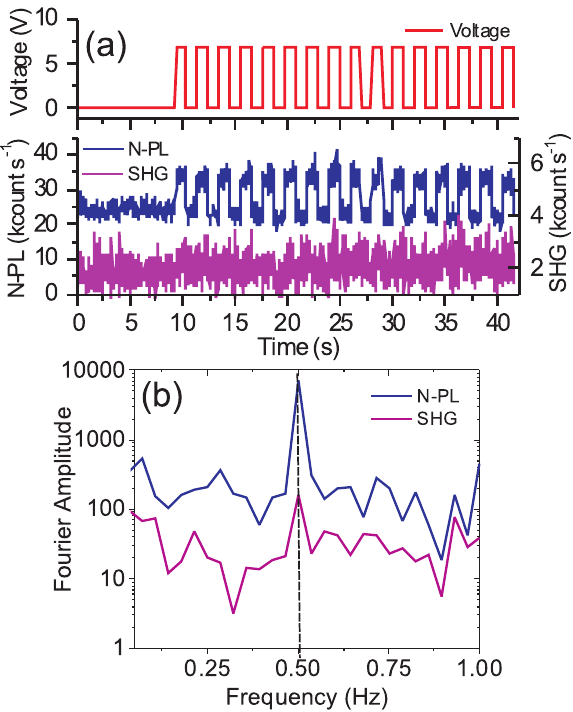}
\caption{(a) Time traces of voltage $V$, N-PL and SHG responses obtained from a 20~nm wide gap antenna. The N-PL rate is modified upon application of 7~V bias while the SHG remains weakly affected. (b) Fourier transform of the modulated signals. The N-PL peak present at the modulation frequency (0.5~Hz) is significantly stronger than the SHG. }
\label{SHG}
\end{figure}

Let us review the different mechanisms that can explain a voltage-sensitive N-PL response. Possible contributors include electrostriction, field-induced morphological reorganization, third-order $\chi^{(3)}$ nonlinearity, and various hot electron phenomena.  Electrostriction is a mechanical deformation of any material upon applying an electrical stress. In the gap geometry considered here, the phenomenon would manifest itself as a variation of gap width due to an elastic deformation of the Au terminals, the supporting SiO$_2$ substrate, or a combination thereof. The electromagnetic response of the gap is essentially dictated by the distance separating the two Au electrodes. Within the range of the gap sizes considered here (tens of nanometers), the local electromagnetic field in the gap is strongly enhanced when near-field coupling between the two metal sides is favored. Because of the intrinsic nonlinear nature of the detected luminescence, the signal is a probe of the local electromagnetic field built up in the gap region~\cite{hecht05sciences,Brida15}. Therefore compressive or tensile in-plane stresses induced by the applied voltage woulds simultaneously increase or reduce the N-PL rate. The electrostrictive strain in the material features a quadratic dependence with the electric field~\cite{Tan96}, therefore its sign does not change direction when the field is reversed and cannot explain the polarity-dependent on-state enhancement or quenching of the N-PL rates observed experimentally in Fig.~\ref{polarity}. We confirmed the absence of an electromechanical deformation of the gap by simultaneously monitoring second-harmonic photons (SHG) produced along side the nonlinear photoluminescence~\cite{Chen81}. The SHG signal was previously reported to be sensitive to the size of the gap in coupled geometries and to subtle variations of the shape of a nanostructure~\cite{Berthelot12b,Martin13}. With the optical antennas studied here, the SHG response is generally weaker than the emitted N-PL. Figure~\ref{SHG} (a) shows the times traces of the N-PL and SHG signals collected from a 20~nm gap (nominal) biased with 7~V square waveform. The N-PL follows the voltage drive while the modulation of the SHG remains very weak upon applying a bias in this unloaded optical gap antenna~\cite{Cai11}. Figure~\ref{SHG}(b) shows the Fourier transform of the varying signals around the modulation frequency (here 0.5~Hz). The graph hints to the presence of a residual response of the SHG to the time-varying applied field, but remains many orders of magnitudes smaller than the N-PL response.  

Another phenomenon that could contribute to a change in the N-PL rate is a modification of the morphology of the optical gap antenna either imposed by the laser~\cite{bouhelier05PRL} or a by a field-induced reorganization of the metal interface~\cite{Olsson18,Emboras16}. Since the modulation amplitude does not significantly change in the time traces presented above, we conclude that the structure under illumination does not undergo an irreversible reshaping and the intensity of the laser remains below the damage threshold~\cite{bouhelier05PRL}. Like second-harmonic generation arising in small Au structures, N-PL is believed to be a surface effect~\cite{boyd86,agreda18}.  Hence, a reversible electric field-induced reorganization of the metal outer boundaries would necessarily affect the strength of both the N-PL and SHG signals. Reordering of the Au outmost atomic layers was found to occur for local field magnitudes in excess of $25\times10^9$~V$\cdot$m$^{-1}$~\cite{Olsson18}, a value two orders of magnitude higher than typical fields applied in our experiment. Even considering an increased mobility of Au adatoms due to a local laser-induced rise of the temperature, it unlikely explains the difference between on-state enhancement and on-state quenching observed for the N-PL. Although noisy, the SHG signal in Fig.~\ref{SHG} confirms that there is no drastic reorganization of the antenna's surface during the experiment.

An additional possible explanation for the voltage modulation of the N-PL is the activation of a Kerr-type nonlinear susceptibility induced by the strong electrostatic field developing in the gap region. The field displaces charges by polarizing the two sides of the gap thereby creating a disturbance in the number density of the electrons. This sheet of polarization, confined to the first atomic layers, can be taken into account by introducing a correction to the complex dielectric function of Au~\cite{Mitra16}. However, like in electrostriction, the corrective term depends quadratically on the electric field. The presence of $\chi^{(3)}$ nonlinearity is thus not able to account for the N-PL sensitivity to the sign of the voltage and the linear-like trend shown in Fig.~\ref{voltagesweep}(c).  

Let us go back to the origin of the nonlinear photoluminescence and the presence of a hot electron distribution. It has been debated that the up-converted emission and its associated nonlinearity may not arise from a multiple absorption events followed by an interband recombination of electron-hole pairs. Instead, it is hypothesized that the nonlinear behavior stems from the emission process.  The out-of-equilibrium electronic distribution generated by the femtosecond light pulses absorbed by the metal radiates~\cite{Haug15}. It is interesting to note that this scenario bears strong similarity with the anomalous light emission observed in electron-fed antennas, where an overbias emission, in other words an upconverted signal, was also understood from the framework  of a heated electron gas~\cite{welland02,Buret2015}. 

\begin{figure}[h!]
\centering
\includegraphics[width=3.25in]{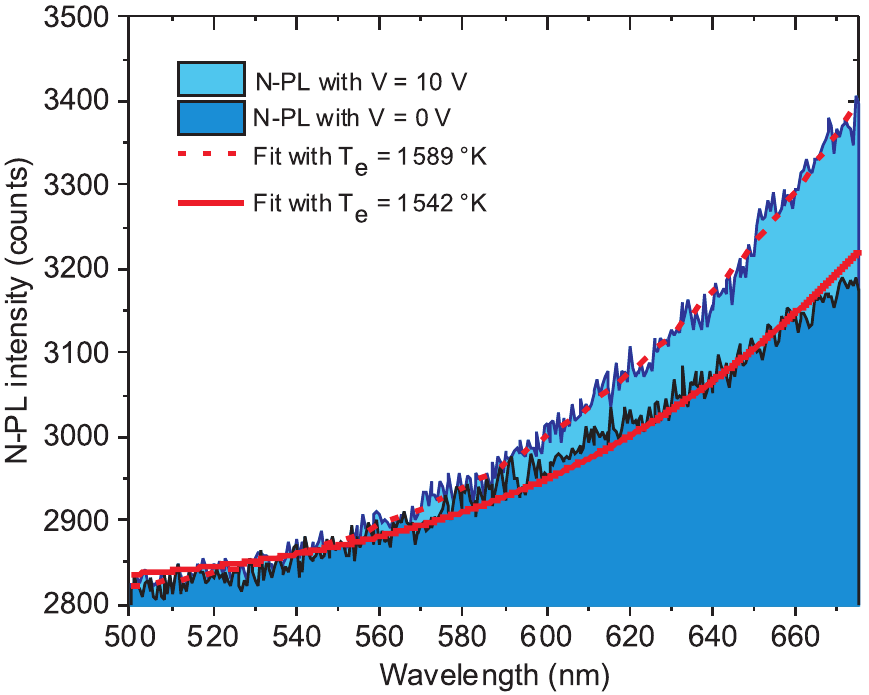}
\caption{Spectra of the nonlinear photoluminescence acquired for two applied voltages $V=0$~V and  $V=10$~V, respectively. The solid lines are fits using Planck's law leading to two different electron temperatures.}
\label{spectra}
\end{figure}

Figure~\ref{spectra} shows typical N-PL spectra emitted by the gap antennas for two biasing conditions  $V=0$~V and  $V=10$~V.  In this example, the bias leads to an on-state enhancement of the N-PL. Both spectra feature a monotonic tail decaying in the visible part of the spectrum and show no evidence of a surface plasmon response~\cite{bouhelier05PRL}. Following the approach consisting at treating the nonlinear photoluminescence as the glowing fate of hot electrons, we tentatively deduce an effective electron temperature using Planck's law of radiation:

\begin{equation}
I_\lambda\propto\frac{2\pi h c^2}{\lambda^5}\frac{1}{e^{\frac{h c}{\lambda k_{\rm B} T_e}}-1}\label{planck}
\end{equation}
  where $I_\lambda$ is the spectral irradiance, $h$ and  $k_{\rm B}$ are Planck's constant and Boltzmann's constant, and $T_e$ is the electron temperature. The solid and dotted lines are fit to the data using Eq.~\ref{planck}. The two fitted curves follow relatively well the spectral dependence of the N-PL considering a temperature of the electron gas at 1542~K when no bias is applied and a raise of  $T_e=1589$~K under a 10~V voltage bias.  Both temperatures are in the same order of magnitude as previously reported values for electronic temperatures reached after pulsed laser excitation~\cite{Gamaly11,Demichel16}

The daunting question to answer is how can a static bias influence the temperature of an out-of-equilibrium electron gas generated by the absorption of a light pulse. 
In a 1998 paper, A.~P.~Kanavin and co-workers derived an analytical expression for $T_e$ when metals are irradiated with femto-second light pulses~\cite{kanavin98}.  They found that  for times $t$ longer than the equilibration of the electron distribution, the electron temperature becomes inversely proportional to the electron density $N_e$ if $T_e$ exceeds a characteristic temperature $T_{\ast}=\sqrt{\varepsilon_{\rm F} T_{\rm L}/ k_{\rm B}}$ where $\varepsilon_{\rm F}$ is the Fermi energy and $T_{\rm L}$ is the phonon temperature. Considering $T_{\rm L}=300$~K and $\varepsilon_{\rm F}=5.53$~eV for Au, $T_{\ast} \sim 4\times 10^3 $~K.  
For the laser power used in the present experiment, the electron temperature after the pulse is expected to be smaller than  $T_{\ast}$. We thus derive in the Sect. 3--5 of the Supporting Information an analytical expression for $T_e$ when $T_e<T_{\ast}$. We find 

\begin{equation}
T_e=\sqrt{\frac{F}{F_0} T_{\rm F} T_{\rm L}}
\label{Te}
\end{equation}
$F$ is the laser fluence, $T_{\rm F}$ is the Fermi temperature, and $F_0$ is a characteristic fluence and writes

\begin{equation}
F_0=\frac{\pi^2}{4}N_e k_{\rm B} \sqrt{\frac{\pi}{6}\frac{T_{\rm F} T_{\rm L}\hbar\tau_p}{m_e}}
\label{F0}
\end{equation}
$\tau_p$ is the pulse duration and $m_e$ is the electron mass. Equation~\ref{Te} shows that if the electron temperature remains below $T_{\ast}$, it no longer depends on $1/N_e$ but on $1/\sqrt{N_e}$. A similar dependence was also found by Agranat and co-workers~\cite{Agranat15}. 

To check the validity of our reasoning, we estimate what should the laser fluence $F$ to reach a maximum electron temperature of about 1500~K. Taking a typical experimental laser power at  $49\times 10^{-6}$~W and the repetition rate of the laser at $80\times 10^{6}$~Hz, the laser pulse energy is $0.61\times 10^{-12}$~J. We find that in order to reach the desired $T_e$, the illumination area should have a diameter of  approximately 450 nm. This value is in very good agreement with the point spread function of the high N.A. objective used in this study. 

In Eq.~\ref{Te}, the only parameter that may be changed upon applying a bias is the local density of electrons present at the surface of the two capacitively coupled electrodes. The influence of carrier concentration on the intensity of linear photoluminescence was already pointed out for organic polymers~\cite{Barbara01}. The static electric field dropped in the gap introduces polarization charges present within the Thomas-Fermi screening length of the metal leading to an accumulation of electrons on the ground side and a depletion of electrons on the positive lead. Using Eq.~\ref{Te}, we find that $\frac{\Delta T_e}{T_e}=\frac{1}{2}\frac{\Delta N_e}{N_e}$.  Assuming an unperturbed electron density at $N_e=5.9\times10^{28}$~m$^{-3}$ for $V_b=0$~V, we evaluate the required change of $N_e$ to account for an increased electron temperature $\Delta T_e=1598-1542=47$~K when switching the bias from $V_b=0$~V to $V_b=10$~V.  We find that  $\Delta T_e$ can be explained by $\Delta N_e=3.59$~nm$^{-3}$. To obtain a surface charge density, we make the assumption that most of the density change is contained within the Thomas-Fermi screening length $L_{\rm TF} $, which is about 60~pm for Au~\cite{Ashcroft-Mermin1968}.
Correspondingly, the change of the electron surface density is then $\Delta_{N_e} \times L_{\rm TF}=3.4\times 10^{-2}$~C$\cdot$m$^{-2}$. 

\begin{figure}[h!]
\centering
\includegraphics[width=3.25in]{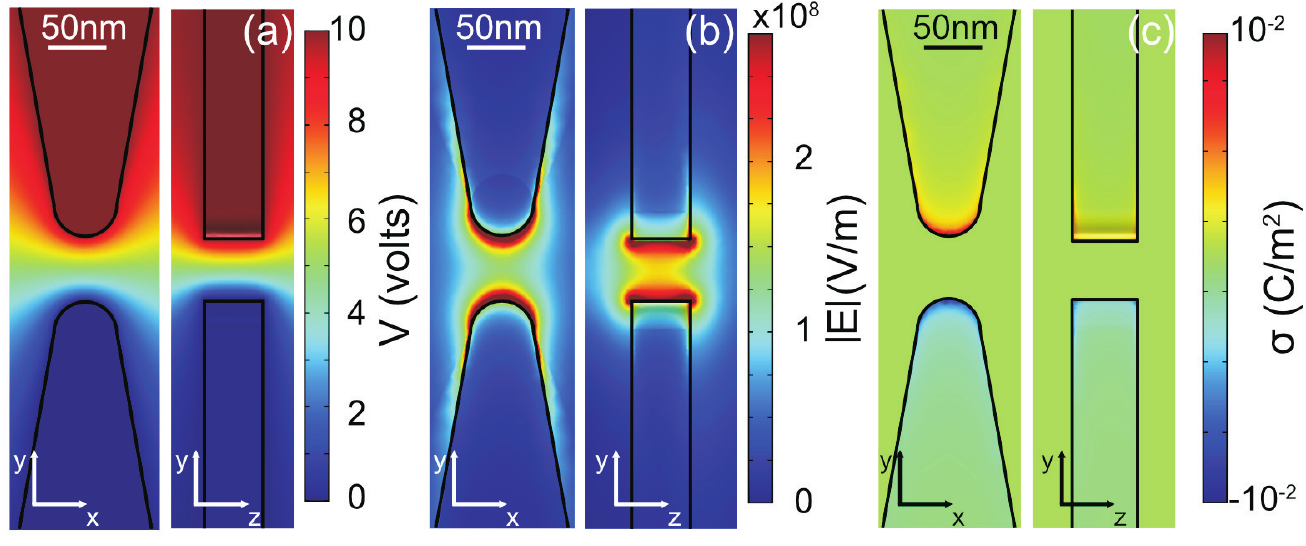}
\caption{ (a) Electrical potential distribution around the gap region pictured from a top view (left frame, ($y,x$) plane) and from a plane normal to the interface (right frame, ($y,z$) plane).  The upper electrode is held at 10~V. (b) Corresponding electric field magnitude.  (c) surface charge density $\sigma (x,y)$ distribution in the gap region. }
\label{simulation}
\end{figure}

In the following, we use a finite element method software (COMSOL)  to obtain the charge distribution in the region of the gap and to compare it to the above figure. The model considers an ideal capacitor with perfectly conductive electrodes, smooth interfaces and a bow-tie shape mimicking the real device. The boundary conditions are as follows. Two gold terminals with a potential difference of 10 V are separated by a 50 nm gap. The gold structures are resting on a silica glass substrate modeled by a cuboid of 1.5 $\mu$m$\times$1.5 $\mu$m$\times$100 nm. The superstrate is air. Figure~\ref{simulation}(a) shows the electrical potential around the gap region. The upper electrode is here held at 10~V. The magnitude of the calculated electric field is about $2.7\times10^8$~V$\cdot$m$^{-1}$ in Fig.~\ref{simulation}(b). The bow-tie geometry leads to a slightly enhanced electric field compared to a capacitor formed by two parallel plates facing each other. Figure~\ref{simulation}(c) displays the corresponding surface charge density $\sigma$. As expected, the largest charge accumulation occurs at the foremost region where the electric field is the strongest.  Under a 10~V bias, the value of $\sigma$ at the interface reaches $\pm 10^{-2}$C$\cdot$m$^{-2}$. The simulated number is in comforting agreement with the modification of the electron density estimated above. These consistencies support us in our scenario: the voltage modulation of the N-PL originates from an electrical control of the electron temperature via a modification of the surface charge density. This also explains why modulation of the N-PL is not observed at the edges of the tapered section of the bow-tie since, under a biased condition, the surface charge density preferentially concentrates at the tip.  Our finding favors the school of thought explaining the nonlinearity of the photoluminescence emission from the radiative decay of a heated electron gas colliding with the interface. Note that the elevated electronic temperature may still result from a multi-photon absorption process of the laser pulse~\cite{Roloff17}. 

\begin{figure}[h!]
\centering
\includegraphics[width=3.25in]{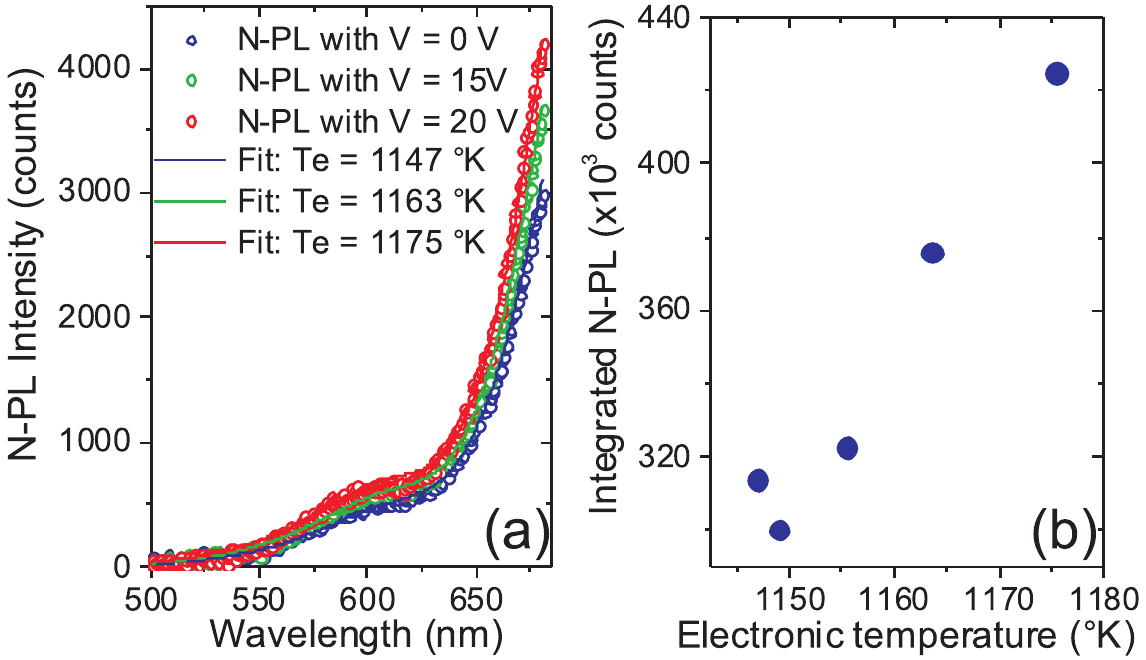}
\caption{ (a) N-PL emission spectrum of optical gap antenna for different biasing conditions. The solid lines are fit to the data using Eq.~\ref{planck} modified by the presence of surface-plasmon like resonance at around 600 nm. (b) Integrated N-PL intensity versus the deduced electronic temperature.}
\label{spectra_vs_Te}
\end{figure}

Since the number of electrons confined at the interface scales in proportion to the applied voltage, the linear trend observed in Fig.~\ref{voltagesweep}(c) is rather odd. The relationship between the N-PL thermal emission is a nonlinear function of $T_e$  (Eq.~\ref{planck}) and $T_e$ is itself inversely proportional to $\sqrt{N_e}$.  We claim that the range of temperatures explored by the voltage activation is small enough so that the trend can be linearized. To support this hypothesis, we investigated the voltage-dependent N-PL emission spectrum of another gap antenna as shown in Fig.~\ref{spectra_vs_Te} (a). The data are fitted again with Planck's law to extract the corresponding electronic temperature. In this set of curves, the spectra (open circles) feature a small resonance in addition to the black-body like tail. Hence, a plasmon contribution in the form of a Gaussian shape is phenomenologically added to Eq.~\ref{planck} to take into account for the shoulder present at around 600 nm. The solid lines are the resulting fits obtained for the different bias contributions.  The fits are satisfactorily following the experimental data points. In Fig.~\ref{spectra_vs_Te} (b) we plot the N-PL intensity, integrated over the entire spectral range, as a function of deduced electronic temperature $T_e$ for all the control voltage values utilized.  The evolution features a quasi linear trend in this  relatively small excursion of the electronic temperature. The apparent linearity confirms our hypothesis that the explored temperature range is too small to observe the nonlinearity  expected from Eq.~\ref{planck} and explains the linear N-PL evolution observed as function of voltage in Fig.~\ref{voltagesweep}(c).

There is one part of the data set shown above that is not completely explained by the proposed mechanism. As shown in Fig.~\ref{pulse}, the modulation is characterized by a combination of a slow and a fast process. Clearly, the charge and the discharge of the gap is dictated by the value of the capacitance, which is here calculated at $C=6.5\times 10^{-18}$ F. Considering a 50~$\Omega$ load, the gap is polarized within a fraction of a femtosecond.  Inevitably, the presence of a millisecond dynamics hints the presence of additional mechanisms not taken into account so far.

The electric fields generated in the gap are fairly strong: the calculation for a 10 V applied voltage shown in Fig.~\ref{simulation}(b) indicates an electric field approaching $3\times10^8$~V$\cdot$m$^{-1}$. Such a large field imposes an electrical stress on the SiO$_2$ substrate, which lead to a wear out of  the oxide and eventually to its dielectric breakdown, when a percolation path forms between the two electrodes~\cite{lombardo05}. The alteration of the insulator layer is associated to the presence of native and stress-induced electron traps in the oxide. 

Among the many different contributors leading to the creation of charge trapping defects and states located at the Au/SiO$_2$ interfaces, a relevant mechanism for the discussion involves the electrical injection of hot holes and hot electrons in the oxide by Fowler-Nordheim tunneling through the triangular potential barrier~\cite{theis84,dimaria89}. In the present experiment, we may discriminate two populations of hot carriers. A first population gains its energy by the lateral field applied across the gap. These carriers injected in the dielectric typically may be trapped by native defects, and may as well generate new trapping sites by releasing hydrogen species that are diffusing toward the counter electrode and producing interface states and electron traps~\cite{Dimaria93,Maes99}. These traps can be charged upon application of the bias by tunneling of electrons, and discharged on removal of the voltage pulse~\cite{Dumin95}. The detrapping lifetime depends on the nature of the centers, their location within the dielectric, the strength and the duration of the electrical stress and a wide range of values spanning $10^{-12}$~s to $10^{3}$~s are reported in the literature~\cite{Balk,vanDriel97,Kang99}. 

A second population of hot electrons, at the origin of the N-PL signal, is created following the absorption of the laser pulse in the metal electrodes and swiftly decays within about a picosecond~\cite{Demichel16}.  This photo-induced out-of-equilibrium electronic distribution drifts with the static electric field whenever $V$ is applied. We hypothesize that the slow part of the nonlinear photoluminescence kinetics is dictated by charging and discharging of field-induced electron traps in the dielectric barrier. In an attempt to verify this claim, we identify the formation of defects in the oxide by monitoring a leakage current transported through the gap between the two electrodes. 
\begin{figure}[h!]
\centering
\includegraphics[width=3.25in]{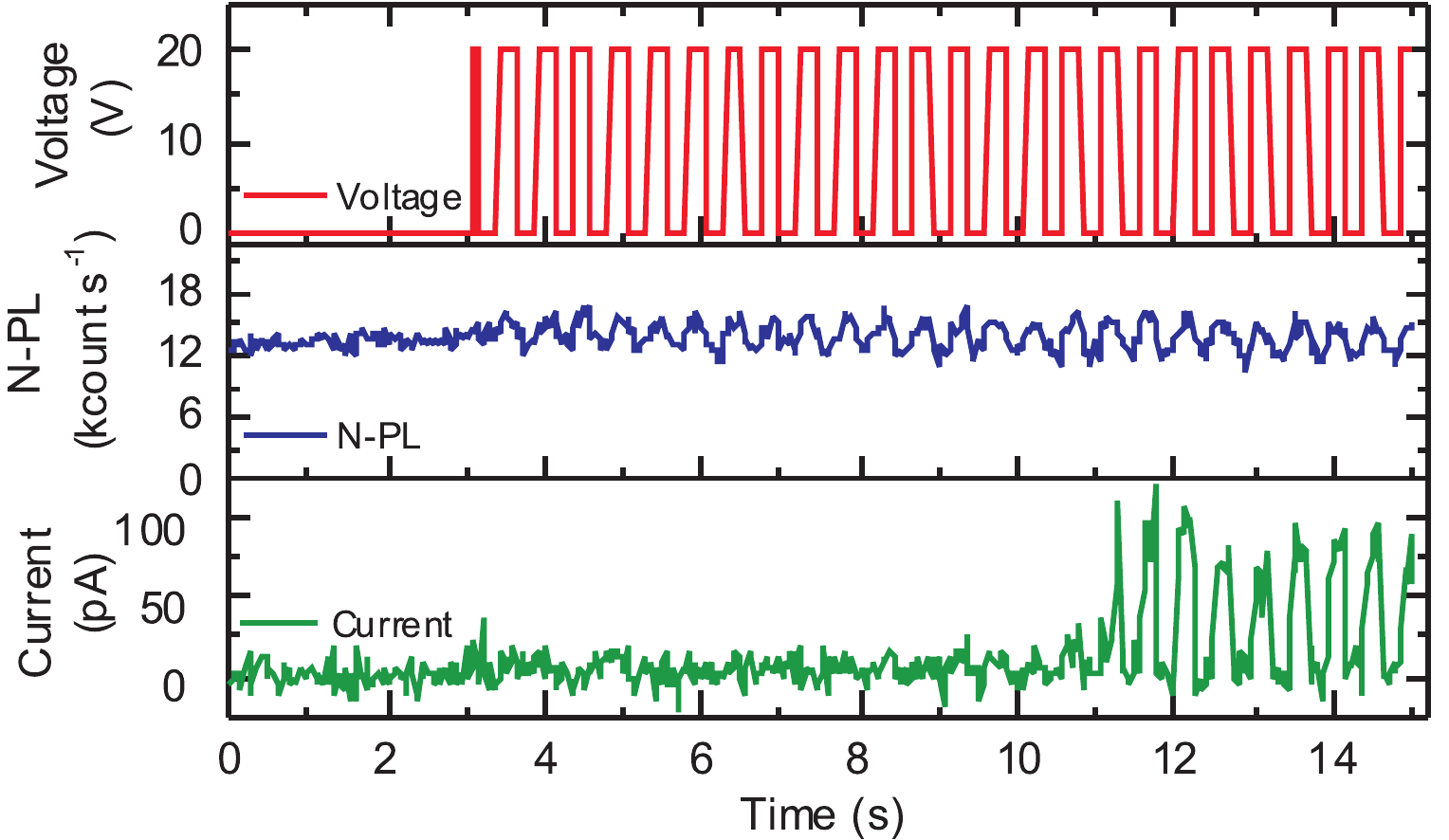}
\caption{Time traces showing the voltage, the N-PL signal and the current. The N-PL responds to the bias with an on-state enhancement. At $t\sim11$~s, an onset of leakage current is measured resulting from the  degradation of the electrically stressed dielectric medium in the gap.}
\label{current}
\end{figure}
In this particular experiment, a gap of $\sim$100 nm is stressed by a 1~s long square voltage  with an amplitude of 20~V. We record simultaneously the voltage, the N-PL signal generated by the positive terminal and the current. The later is detected by introducing a transimpedance amplifier (Femto GmbH, DLPCA-200) in the electrical circuit with a gain of $10^9$~V/A.  The results are shown in Fig.~\ref{current}. At $t=3$~s, the voltage is turned on. Accordingly, the N-PL responds to the voltage by an on-state enhancement. During the first 10~s, no current is detected, as expected from this rather large gap. However, at $t\sim11$~s, we observe a clear signature of a leakage current flowing in the dielectric whenever the bias is applied. These delayed current signal is indicative of build up of defects within the oxide, which eventually get electrically connected and contribute to the wear out of the oxide. Figure~\ref{current} unambiguously demonstrates that the repetitive electrical stress applied to the gap antenna leads to the creation of traps. We note that whatever the nature of the electron pathway through the oxide, it does not affect the modulation ratio of the N-PL. Since N-PL is generated at the Au/SiO$_2$ interface, trapping defects generated deeper in the oxide gap are unlikely to affect the temperature of the photo-induced hot electron population via a change of $N_e$ at the metal outer boundary. We tentatively tried to remove the effect of the substrate by realizing suspended Au structures. However, fabrication difficulties prevented us to produce  tapered bridges stable enough to withstand the optical and the electrical stresses. Other type of amorphous dielectrics, such as HfO$_2$, will also feature electrical-induced wear-out. We discarded crystalline quartz substrate because of its piezoelectric and triboluminescence responses.

We can draw two sets of conclusions from the series of experiments presented here. The first set has a technological character. We have experimentally demonstrated that the nonlinear response of electrically connected optical gap antennas can be externally amplified or attenuated by a control voltage. The modulation occurs without the addition of nonlinear polymers and relies on the intrinsic metal nonlinearity.  This type of electrical control, demonstrated here with Au nonlinear photoluminescence (and to a lesser extend with the second-harmonic generation), opens new avenues for actuating and reconfiguring more complex device architectures such as plasmonic logic gates for instance~\cite{Kumar18}. The second set of conclusions addresses the physical origin of the voltage-controlled nonlinear photoluminescence yield. We propose a scenario by which the local electron temperature at the metal interface is modified by a sheet of polarization charges created by the static electric field. By changing the surface density of electrons to about  $10^{-2}$C$\cdot$m$^{-2}$, the temperature of the electronic distribution can be adjusted by $\sim$ 50~K. This approach can be combined with designs optimized for generating non-uniform electron temperature~\cite{Wiederrecht15,Zayats19} in order to bring additional agility. The present finding  focuses on the response of a bow-tie shape, combining a large N-PL yield overlapping spatially with the region where the electrical potential is dropped. These conditions may be found in other geometries.


The work was made possible through the Indo-French Centre for the Promotion of Advanced Research, project No. 5504-3, the project APEX funded by the Conseil R\'egional de Bourgogne Franche-Comt\'e, the European Regional Development Fund (ERDF), and the CNRS/RFBR collaborative research program number 1493 (Grant RFBR-17-58-150007). Device fabrication was performed in the technological platform ARCEN Carnot with the support of the R\'egion de Bourgogne Franche-Comt\'e and the D\'el\'egation R\'egionale \`a la Recherche et \`a la Technologie (DRRT).

{\bf Supporting Information Available:} fabrication protocol, power dependence of the nonlinear emission, simplified model of metal heating by a short laser pulse, perturbation theory, peak electron temperature at the contact surface. This material is available free of charge via the Internet at http://pubs.acs.org


\begin{mcitethebibliography}{59}
\providecommand*\natexlab[1]{#1}
\providecommand*\mciteSetBstSublistMode[1]{}
\providecommand*\mciteSetBstMaxWidthForm[2]{}
\providecommand*\mciteBstWouldAddEndPuncttrue
  {\def\EndOfBibitem{\unskip.}}
\providecommand*\mciteBstWouldAddEndPunctfalse
  {\let\EndOfBibitem\relax}
\providecommand*\mciteSetBstMidEndSepPunct[3]{}
\providecommand*\mciteSetBstSublistLabelBeginEnd[3]{}
\providecommand*\EndOfBibitem{}
\mciteSetBstSublistMode{f}
\mciteSetBstMaxWidthForm{subitem}{(\alph{mcitesubitemcount})}
\mciteSetBstSublistLabelBeginEnd
  {\mcitemaxwidthsubitemform\space}
  {\relax}
  {\relax}

\bibitem[Kauranen and Zayats(2012)Kauranen, and Zayats]{Kauranen12}
Kauranen,~M.; Zayats,~A.~V. Nonlinear plasmonics. \emph{Nat. Photonics}
  \textbf{2012}, \emph{6}, 737--748\relax
\mciteBstWouldAddEndPuncttrue
\mciteSetBstMidEndSepPunct{\mcitedefaultmidpunct}
{\mcitedefaultendpunct}{\mcitedefaultseppunct}\relax
\EndOfBibitem
\bibitem[Gallinet \latin{et~al.}(2015)Gallinet, Butet, and Martin]{Martin15}
Gallinet,~B.; Butet,~J.; Martin,~O. J.~F. Numerical methods for nanophotonics:
  standard problems and future challenges. \emph{Las. \& Phot. Rev.}
  \textbf{2015}, \emph{9}, 577--603\relax
\mciteBstWouldAddEndPuncttrue
\mciteSetBstMidEndSepPunct{\mcitedefaultmidpunct}
{\mcitedefaultendpunct}{\mcitedefaultseppunct}\relax
\EndOfBibitem
\bibitem[Celebrano \latin{et~al.}(2015)Celebrano, Wu, Baselli, Gro{\ss}mann,
  Biagioni, Locatelli, De~Angelis, Cerullo, Osellame, Hecht, Du{\`o}, Ciccacci,
  and Finazzi]{Celebrano15}
Celebrano,~M.; Wu,~X.; Baselli,~M.; Gro{\ss}mann,~S.; Biagioni,~P.;
  Locatelli,~A.; De~Angelis,~C.; Cerullo,~G.; Osellame,~R.; Hecht,~B.;
  Du{\`o},~L.; Ciccacci,~F.; Finazzi,~M. Mode matching in multiresonant
  plasmonic nanoantennas for enhanced second harmonic generation. \emph{Nature
  Nanotech.} \textbf{2015}, \emph{10}, 412--417\relax
\mciteBstWouldAddEndPuncttrue
\mciteSetBstMidEndSepPunct{\mcitedefaultmidpunct}
{\mcitedefaultendpunct}{\mcitedefaultseppunct}\relax
\EndOfBibitem
\bibitem[Hache \latin{et~al.}(1985)Hache, Ricard, Flytzanis, and
  Kreibig]{Hache85}
Hache,~F.; Ricard,~D.; Flytzanis,~C.; Kreibig,~U. The Optical Kerr Effect in
  Small Metal Particles and Metals Colloids: the Case of Gold. \emph{Appl.
  Phys. A} \textbf{1985}, \emph{47}, 347--357\relax
\mciteBstWouldAddEndPuncttrue
\mciteSetBstMidEndSepPunct{\mcitedefaultmidpunct}
{\mcitedefaultendpunct}{\mcitedefaultseppunct}\relax
\EndOfBibitem
\bibitem[Boyd \latin{et~al.}(1986)Boyd, Yu, and Shen]{boyd86}
Boyd,~G.~T.; Yu,~Z.~H.; Shen,~Y.~R. Photoinduced luminescence from the noble
  metals and its enhancement on roughned surface. \emph{Phys. Rev. B}
  \textbf{1986}, \emph{33}, 7923--7936\relax
\mciteBstWouldAddEndPuncttrue
\mciteSetBstMidEndSepPunct{\mcitedefaultmidpunct}
{\mcitedefaultendpunct}{\mcitedefaultseppunct}\relax
\EndOfBibitem
\bibitem[Voisin \latin{et~al.}(2004)Voisin, Christofilos, Loukakos, Del~Fatti,
  Vall\'ee, Lerm\'e, Gaudry, Cottancin, Pellarin, and Broyer]{Voisin04}
Voisin,~C.; Christofilos,~D.; Loukakos,~P.~A.; Del~Fatti,~N.; Vall\'ee,~F.;
  Lerm\'e,~J.; Gaudry,~M.; Cottancin,~E.; Pellarin,~M.; Broyer,~M. Ultrafast
  electron-electron scattering and energy exchanges in noble-metal
  nanoparticles. \emph{Phys. Rev. B} \textbf{2004}, \emph{69}, 195416\relax
\mciteBstWouldAddEndPuncttrue
\mciteSetBstMidEndSepPunct{\mcitedefaultmidpunct}
{\mcitedefaultendpunct}{\mcitedefaultseppunct}\relax
\EndOfBibitem
\bibitem[Hou \latin{et~al.}(2018)Hou, Djellali, and Palpant]{Palpant18}
Hou,~X.; Djellali,~N.; Palpant,~B. Absorption of Ultrashort Laser Pulses by
  Plasmonic Nanoparticles: Not Necessarily What You Might Think. \emph{ACS
  Phot.} \textbf{2018}, \emph{5}, 3856--3863\relax
\mciteBstWouldAddEndPuncttrue
\mciteSetBstMidEndSepPunct{\mcitedefaultmidpunct}
{\mcitedefaultendpunct}{\mcitedefaultseppunct}\relax
\EndOfBibitem
\bibitem[Berthelot \latin{et~al.}(2012)Berthelot, Bachelier, Song, Rai, {Colas
  des Francs}, Dereux, and Bouhelier]{Berthelot12b}
Berthelot,~J.; Bachelier,~G.; Song,~M.; Rai,~P.; {Colas des Francs},~G.;
  Dereux,~A.; Bouhelier,~A. Silencing and enhancement of second-harmonic
  generation in optical gap antennas. \emph{Opt. Express} \textbf{2012},
  \emph{20}, 10498--10508\relax
\mciteBstWouldAddEndPuncttrue
\mciteSetBstMidEndSepPunct{\mcitedefaultmidpunct}
{\mcitedefaultendpunct}{\mcitedefaultseppunct}\relax
\EndOfBibitem
\bibitem[Viarbitskaya \latin{et~al.}(2013)Viarbitskaya, Teulle, Marty, Sharma,
  Girard, Arbouet, and Dujardin]{Viarbitskaya2013}
Viarbitskaya,~S.; Teulle,~A.; Marty,~R.; Sharma,~J.; Girard,~C.; Arbouet,~A.;
  Dujardin,~E. Tailoring and imaging the plasmonic local density of states in
  crystalline nanoprisms. \emph{Nature Mat.} \textbf{2013}, \emph{12},
  426--432\relax
\mciteBstWouldAddEndPuncttrue
\mciteSetBstMidEndSepPunct{\mcitedefaultmidpunct}
{\mcitedefaultendpunct}{\mcitedefaultseppunct}\relax
\EndOfBibitem
\bibitem[Comin and Hartschuh(2018)Comin, and Hartschuh]{Comin18}
Comin,~A.; Hartschuh,~A. Efficient optimization of SHG hotspot switching in
  plasmonic nanoantennas using phase-shaped laser pulses controlled by neural
  networks. \emph{Opt. Express} \textbf{2018}, \emph{26}, 33678--33686\relax
\mciteBstWouldAddEndPuncttrue
\mciteSetBstMidEndSepPunct{\mcitedefaultmidpunct}
{\mcitedefaultendpunct}{\mcitedefaultseppunct}\relax
\EndOfBibitem
\bibitem[Remesh \latin{et~al.}(2018)Remesh, St\"uhrenberg, Saemisch, Accanto,
  and van Hulst]{VanHulst18}
Remesh,~V.; St\"uhrenberg,~M.; Saemisch,~L.; Accanto,~N.; van Hulst,~N.~F.
  Phase control of plasmon enhanced two-photon photoluminescence in resonant
  gold nanoantennas. \emph{Appl.Phys.Lett.} \textbf{2018}, \emph{113},
  211101\relax
\mciteBstWouldAddEndPuncttrue
\mciteSetBstMidEndSepPunct{\mcitedefaultmidpunct}
{\mcitedefaultendpunct}{\mcitedefaultseppunct}\relax
\EndOfBibitem
\bibitem[Berline \latin{et~al.}(2008)Berline, Fiorini-Debuisschert, Royal,
  Douillard, and Charra]{Fiorini08}
Berline,~I.; Fiorini-Debuisschert,~C.; Royal,~C.; Douillard,~L.; Charra,~F.
  Molecular second harmonic generation induced at a metallic tip. \emph{J.
  Appl. Phys.} \textbf{2008}, \emph{104}, 103113\relax
\mciteBstWouldAddEndPuncttrue
\mciteSetBstMidEndSepPunct{\mcitedefaultmidpunct}
{\mcitedefaultendpunct}{\mcitedefaultseppunct}\relax
\EndOfBibitem
\bibitem[Cai \latin{et~al.}(2011)Cai, Vasudev, and Brongersma]{Cai11}
Cai,~W.; Vasudev,~A.~P.; Brongersma,~M.~L. Electrically Controlled Nonlinear
  Generation of Light with Plasmonics. \emph{Science} \textbf{2011},
  \emph{333}, 1720--1723\relax
\mciteBstWouldAddEndPuncttrue
\mciteSetBstMidEndSepPunct{\mcitedefaultmidpunct}
{\mcitedefaultendpunct}{\mcitedefaultseppunct}\relax
\EndOfBibitem
\bibitem[Wokaun \latin{et~al.}(1981)Wokaun, Bergman, Heritage, Glass, Liao, and
  Olson]{Wokaum81}
Wokaun,~A.; Bergman,~J.~G.; Heritage,~J.~P.; Glass,~A.~M.; Liao,~P.~F.;
  Olson,~D.~H. Surface second-harmonic generation from metal island films and
  microlithographic structures. \emph{Phys. Rev. B} \textbf{1981}, \emph{24},
  849--856\relax
\mciteBstWouldAddEndPuncttrue
\mciteSetBstMidEndSepPunct{\mcitedefaultmidpunct}
{\mcitedefaultendpunct}{\mcitedefaultseppunct}\relax
\EndOfBibitem
\bibitem[Bachelier \latin{et~al.}(2010)Bachelier, Butet, Russier-Antoine,
  Jonin, Benichou, and Brevet]{Bachelier10}
Bachelier,~G.; Butet,~J.; Russier-Antoine,~I.; Jonin,~C.; Benichou,~E.;
  Brevet,~P.-F. Origin of optical second-harmonic generation in spherical gold
  nanoparticles: Local surface and nonlocal bulk contributions. \emph{Phys.
  Rev. B} \textbf{2010}, \emph{82}, 235403\relax
\mciteBstWouldAddEndPuncttrue
\mciteSetBstMidEndSepPunct{\mcitedefaultmidpunct}
{\mcitedefaultendpunct}{\mcitedefaultseppunct}\relax
\EndOfBibitem
\bibitem[Beversluis \latin{et~al.}(2003)Beversluis, Bouhelier, and
  Novotny]{Beversluis03}
Beversluis,~M.~R.; Bouhelier,~A.; Novotny,~L. Continuum generation from single
  gold nanostructures through near-field mediated intraband transitions.
  \emph{Phys. Rev. B} \textbf{2003}, \emph{68}, 115433\relax
\mciteBstWouldAddEndPuncttrue
\mciteSetBstMidEndSepPunct{\mcitedefaultmidpunct}
{\mcitedefaultendpunct}{\mcitedefaultseppunct}\relax
\EndOfBibitem
\bibitem[Imura \latin{et~al.}(2004)Imura, Nagahara, and Okamoto]{Imura04}
Imura,~K.; Nagahara,~T.; Okamoto,~H. Imaging of Surface Plasmon and Ultrafast
  Dynamics in Gold Nanorods by Near-Field Microscopy. \emph{J. Phys. Chem. B}
  \textbf{2004}, \emph{108}, 16344\relax
\mciteBstWouldAddEndPuncttrue
\mciteSetBstMidEndSepPunct{\mcitedefaultmidpunct}
{\mcitedefaultendpunct}{\mcitedefaultseppunct}\relax
\EndOfBibitem
\bibitem[Biagioni \latin{et~al.}(2009)Biagioni, Celebrano, Savoini, Grancini,
  Brida, M\'at\'efi-Tempfli, M\'at\'efi-Tempfli, Du\`o, Hecht, Cerullo, and
  Finazzi]{Biagoni09}
Biagioni,~P.; Celebrano,~M.; Savoini,~M.; Grancini,~G.; Brida,~D.;
  M\'at\'efi-Tempfli,~S.; M\'at\'efi-Tempfli,~M.; Du\`o,~L.; Hecht,~B.;
  Cerullo,~G.; Finazzi,~M. Dependence of the two-photon photoluminescence yield
  of gold nanostructures on the laser pulse duration. \emph{Phys. Rev. B}
  \textbf{2009}, \emph{80}, 045411\relax
\mciteBstWouldAddEndPuncttrue
\mciteSetBstMidEndSepPunct{\mcitedefaultmidpunct}
{\mcitedefaultendpunct}{\mcitedefaultseppunct}\relax
\EndOfBibitem
\bibitem[Chen \latin{et~al.}(2014)Chen, Lin, Lee, Li, Chang, and
  Huang]{Huang14}
Chen,~W.-L.; Lin,~F.-C.; Lee,~Y.-Y.; Li,~F.-C.; Chang,~Y.-M.; Huang,~J.-S. The
  Modulation Effect of Transverse, Antibonding, and Higher-Order Longitudinal
  Modes on the Two-Photon Photoluminescence of Gold Plasmonic Nanoantennas.
  \emph{ACS Nano} \textbf{2014}, \emph{8}, 9053\relax
\mciteBstWouldAddEndPuncttrue
\mciteSetBstMidEndSepPunct{\mcitedefaultmidpunct}
{\mcitedefaultendpunct}{\mcitedefaultseppunct}\relax
\EndOfBibitem
\bibitem[Hugall and Baumberg(2015)Hugall, and Baumberg]{Baumberg15}
Hugall,~J.~T.; Baumberg,~J.~J. Demonstrating Photoluminescence from Au is
  Electronic InelasticLight Scattering of a Plasmonic Metal: The Origin of SERS
  Backgrounds. \emph{Nano Lett.} \textbf{2015}, \emph{4}, 2600--2604\relax
\mciteBstWouldAddEndPuncttrue
\mciteSetBstMidEndSepPunct{\mcitedefaultmidpunct}
{\mcitedefaultendpunct}{\mcitedefaultseppunct}\relax
\EndOfBibitem
\bibitem[Chen \latin{et~al.}(2018)Chen, Krasavin, Ginzburg, Zayats, Pullerits,
  and Karki]{Karki18}
Chen,~J.; Krasavin,~A.; Ginzburg,~P.; Zayats,~A.~V.; Pullerits,~T.;
  Karki,~K.~J. Evidence of High-Order Nonlinearities in Supercontinuum
  White-Light Generation from a Gold Nanofilm. \emph{ACS Phot.} \textbf{2018},
  \emph{5}, 1927--1932\relax
\mciteBstWouldAddEndPuncttrue
\mciteSetBstMidEndSepPunct{\mcitedefaultmidpunct}
{\mcitedefaultendpunct}{\mcitedefaultseppunct}\relax
\EndOfBibitem
\bibitem[Haug \latin{et~al.}(2015)Haug, Klemm, Bange, and Lupton]{Haug15}
Haug,~T.; Klemm,~P.; Bange,~S.; Lupton,~J.~M. Hot-Electron Intraband
  Luminescence from Single Hot Spots in Noble-Metal Nanoparticle Films.
  \emph{Phys. Rev. Lett.} \textbf{2015}, \emph{115}, 067403\relax
\mciteBstWouldAddEndPuncttrue
\mciteSetBstMidEndSepPunct{\mcitedefaultmidpunct}
{\mcitedefaultendpunct}{\mcitedefaultseppunct}\relax
\EndOfBibitem
\bibitem[Brongersma \latin{et~al.}(2015)Brongersma, Halas, and
  Nordlander]{Nordlander15}
Brongersma,~M.~L.; Halas,~N.~J.; Nordlander,~P. Plasmon-induced hot carrier
  science and technology. \emph{Nat. Nanotechnol.} \textbf{2015}, \emph{10},
  25--34\relax
\mciteBstWouldAddEndPuncttrue
\mciteSetBstMidEndSepPunct{\mcitedefaultmidpunct}
{\mcitedefaultendpunct}{\mcitedefaultseppunct}\relax
\EndOfBibitem
\bibitem[Saavedra \latin{et~al.}(2016)Saavedra, Asenjo-Garcia, and Garc\'ia~de
  Abajo]{DeAbajo16}
Saavedra,~J.; Asenjo-Garcia,~A.; Garc\'ia~de Abajo,~F.~J. Hot-electron dynamics
  and thermalization in small metallic nanoparticles. \emph{ACS Phot.}
  \textbf{2016}, \emph{3}, 1637--1646\relax
\mciteBstWouldAddEndPuncttrue
\mciteSetBstMidEndSepPunct{\mcitedefaultmidpunct}
{\mcitedefaultendpunct}{\mcitedefaultseppunct}\relax
\EndOfBibitem
\bibitem[Viarbitskaya \latin{et~al.}(2015)Viarbitskaya, Cuche, Teulle, Sharma,
  Girard, Arbouet, and Dujardin]{Dujardin15}
Viarbitskaya,~S.; Cuche,~A.; Teulle,~A.; Sharma,~J.; Girard,~C.; Arbouet,~A.;
  Dujardin,~E. Plasmonic Hot Printing in Gold Nanoprisms. \emph{ACS Phot.}
  \textbf{2015}, \emph{2}, 744--751\relax
\mciteBstWouldAddEndPuncttrue
\mciteSetBstMidEndSepPunct{\mcitedefaultmidpunct}
{\mcitedefaultendpunct}{\mcitedefaultseppunct}\relax
\EndOfBibitem
\bibitem[Demichel \latin{et~al.}(2016)Demichel, Petit, Viarbitskaya,
  M{\'e}jard, de~Fornel, Hertz, Billard, Bouhelier, and Cluzel]{Demichel16}
Demichel,~O.; Petit,~M.; Viarbitskaya,~S.; M{\'e}jard,~R.; de~Fornel,~F.;
  Hertz,~E.; Billard,~F.; Bouhelier,~A.; Cluzel,~B. Dynamics, Efficiency, and
  Energy Distribution of Nonlinear Plasmon-Assisted Generation of Hot Carriers.
  \emph{ACS Phot.} \textbf{2016}, \emph{3}, 791--795\relax
\mciteBstWouldAddEndPuncttrue
\mciteSetBstMidEndSepPunct{\mcitedefaultmidpunct}
{\mcitedefaultendpunct}{\mcitedefaultseppunct}\relax
\EndOfBibitem
\bibitem[Bouhelier \latin{et~al.}(2005)Bouhelier, Bachelot, Lerondel,
  Kostcheev, Royer, and Wiederrecht]{bouhelier05PRL}
Bouhelier,~A.; Bachelot,~R.; Lerondel,~G.; Kostcheev,~S.; Royer,~P.;
  Wiederrecht,~G.~P. Surface Plasmon Characteristics of Tunable
  Photoluminescence in Single Gold Nanorods. \emph{Phys. Rev. Lett.}
  \textbf{2005}, \emph{95}, 267405\relax
\mciteBstWouldAddEndPuncttrue
\mciteSetBstMidEndSepPunct{\mcitedefaultmidpunct}
{\mcitedefaultendpunct}{\mcitedefaultseppunct}\relax
\EndOfBibitem
\bibitem[M\"uhlschlegel \latin{et~al.}(2005)M\"uhlschlegel, Eisler, Martin,
  Hecht, and Pohl]{hecht05sciences}
M\"uhlschlegel,~P.; Eisler,~H.-J.; Martin,~O. J.~F.; Hecht,~B.; Pohl,~D.~W.
  Resonant Optical Antennas. \emph{Science} \textbf{2005}, \emph{308},
  1607--1609\relax
\mciteBstWouldAddEndPuncttrue
\mciteSetBstMidEndSepPunct{\mcitedefaultmidpunct}
{\mcitedefaultendpunct}{\mcitedefaultseppunct}\relax
\EndOfBibitem
\bibitem[Knittel \latin{et~al.}(2015)Knittel, Fischer, de~Roo, Mecking,
  Leitenstorfer, and Brida]{Brida15}
Knittel,~V.; Fischer,~M.~P.; de~Roo,~T.; Mecking,~S.; Leitenstorfer,~A.;
  Brida,~D. Nonlinear photoluminescence spectrum of single gold nanostructures.
  \emph{ACS Nano} \textbf{2015}, \emph{9}, 894--900\relax
\mciteBstWouldAddEndPuncttrue
\mciteSetBstMidEndSepPunct{\mcitedefaultmidpunct}
{\mcitedefaultendpunct}{\mcitedefaultseppunct}\relax
\EndOfBibitem
\bibitem[Cuche \latin{et~al.}(2014)Cuche, Viarbitskaya, Sharma, Arbouet,
  Girard, and Dujardin]{cuche14}
Cuche,~A.; Viarbitskaya,~S.; Sharma,~J.; Arbouet,~A.; Girard,~C.; Dujardin,~E.
  Modal engineering of Surface Plasmons in apertured Au Nanoprisms. \emph{Sci.
  Rep.} \textbf{2014}, \emph{5}, 16635\relax
\mciteBstWouldAddEndPuncttrue
\mciteSetBstMidEndSepPunct{\mcitedefaultmidpunct}
{\mcitedefaultendpunct}{\mcitedefaultseppunct}\relax
\EndOfBibitem
\bibitem[Al\`{u} and Engheta(2008)Al\`{u}, and Engheta]{engheta08}
Al\`{u},~A.; Engheta,~N. Input Impedance, Nanocircuit Loading, and Radiation
  Tuning of Optical Nanoantennas. \emph{Phy. Rev. Lett.} \textbf{2008},
  \emph{101}, 043901\relax
\mciteBstWouldAddEndPuncttrue
\mciteSetBstMidEndSepPunct{\mcitedefaultmidpunct}
{\mcitedefaultendpunct}{\mcitedefaultseppunct}\relax
\EndOfBibitem
\bibitem[Tan and Arndt(1996)Tan, and Arndt]{Tan96}
Tan,~C.; Arndt,~J. Measurement of piezoelectricity in quartz and
  electrostriction in SiO2 glass by interferometric method. \emph{Phys. B:
  Cond. Mat.} \textbf{1996}, \emph{225}, 202 -- 206\relax
\mciteBstWouldAddEndPuncttrue
\mciteSetBstMidEndSepPunct{\mcitedefaultmidpunct}
{\mcitedefaultendpunct}{\mcitedefaultseppunct}\relax
\EndOfBibitem
\bibitem[Chen \latin{et~al.}(1981)Chen, {de Castro}, and Chen]{Chen81}
Chen,~C.~K.; {de Castro},~A. R.~B.; Chen,~Y.~R. Surface-Enhanced Second
  Harmonic Generation. \emph{Phys. Rev. Lett.} \textbf{1981}, \emph{46},
  145--148\relax
\mciteBstWouldAddEndPuncttrue
\mciteSetBstMidEndSepPunct{\mcitedefaultmidpunct}
{\mcitedefaultendpunct}{\mcitedefaultseppunct}\relax
\EndOfBibitem
\bibitem[Butet \latin{et~al.}(2013)Butet, Thyagarajan, and Martin]{Martin13}
Butet,~J.; Thyagarajan,~K.; Martin,~O. J.~F. Ultrasensitive Optical Shape
  Characterization of Gold Nanoantennas Using Second Harmonic Generation.
  \emph{Nano Lett.} \textbf{2013}, \emph{13}, 1787--1792\relax
\mciteBstWouldAddEndPuncttrue
\mciteSetBstMidEndSepPunct{\mcitedefaultmidpunct}
{\mcitedefaultendpunct}{\mcitedefaultseppunct}\relax
\EndOfBibitem
\bibitem[de~Knoop \latin{et~al.}(2018)de~Knoop, Juhani~Kuisma, L\"ofgren,
  Lodewijks, Thuvander, Erhart, Dmitriev, and Olsson]{Olsson18}
de~Knoop,~L.; Juhani~Kuisma,~M.; L\"ofgren,~J.; Lodewijks,~K.; Thuvander,~M.;
  Erhart,~P.; Dmitriev,~A.; Olsson,~E. Electric-field-controlled reversible
  order-disorder switching of a metal tip surface. \emph{Phys. Rev. Mat.}
  \textbf{2018}, \emph{2}, 085006\relax
\mciteBstWouldAddEndPuncttrue
\mciteSetBstMidEndSepPunct{\mcitedefaultmidpunct}
{\mcitedefaultendpunct}{\mcitedefaultseppunct}\relax
\EndOfBibitem
\bibitem[Emboras \latin{et~al.}(2016)Emboras, Niegemann, Ma, Haffner, Pedersen,
  Luisier, Hafner, Schimmel, and Leuthold]{Emboras16}
Emboras,~A.; Niegemann,~J.; Ma,~P.; Haffner,~C.; Pedersen,~A.; Luisier,~M.;
  Hafner,~C.; Schimmel,~T.; Leuthold,~J. Atomic Scale Plasmonic Switch.
  \emph{Nano Lett.} \textbf{2016}, \emph{16}, 709--714\relax
\mciteBstWouldAddEndPuncttrue
\mciteSetBstMidEndSepPunct{\mcitedefaultmidpunct}
{\mcitedefaultendpunct}{\mcitedefaultseppunct}\relax
\EndOfBibitem
\bibitem[Agreda \latin{et~al.}(2019)Agreda, Sharma, Viarbitskaya, Hernandez,
  Cluzel, Demichel, Weeber, Colas~des Francs, Kumar, and Bouhelier]{agreda18}
Agreda,~A.; Sharma,~D.~K.; Viarbitskaya,~S.; Hernandez,~R.; Cluzel,~B.;
  Demichel,~O.; Weeber,~J.-C.; Colas~des Francs,~G.; Kumar,~G.~P.;
  Bouhelier,~A. Spatial Distribution of the Nonlinear Photoluminescence in Au
  Nanowires. \emph{ACS Phot.} \textbf{2019}, \emph{6}, 1240--1247\relax
\mciteBstWouldAddEndPuncttrue
\mciteSetBstMidEndSepPunct{\mcitedefaultmidpunct}
{\mcitedefaultendpunct}{\mcitedefaultseppunct}\relax
\EndOfBibitem
\bibitem[Kalathingal \latin{et~al.}(2016)Kalathingal, Dawson, and
  Mitra]{Mitra16}
Kalathingal,~V.; Dawson,~P.; Mitra,~J. Scanning tunneling microscope light
  emission: Effect of the strong dc field on junction plasmons. \emph{Phys.
  Rev. B} \textbf{2016}, \emph{94}, 035443\relax
\mciteBstWouldAddEndPuncttrue
\mciteSetBstMidEndSepPunct{\mcitedefaultmidpunct}
{\mcitedefaultendpunct}{\mcitedefaultseppunct}\relax
\EndOfBibitem
\bibitem[Downes \latin{et~al.}(2002)Downes, Dumas, and Welland]{welland02}
Downes,~A.; Dumas,~P.; Welland,~M.~E. Measurement of high electron temperature
  in single atom metal point contacts by light emission. \emph{Appl. Phys.
  Lett.} \textbf{2002}, \emph{81}, 1252--1254\relax
\mciteBstWouldAddEndPuncttrue
\mciteSetBstMidEndSepPunct{\mcitedefaultmidpunct}
{\mcitedefaultendpunct}{\mcitedefaultseppunct}\relax
\EndOfBibitem
\bibitem[Buret \latin{et~al.}(2015)Buret, Uskov, Dellinger, Cazier,
  Mennemanteuil, Berthelot, Smetanin, Protsenko, {Colas-des-Francs}, and
  Bouhelier]{Buret2015}
Buret,~M.; Uskov,~A.~V.; Dellinger,~J.; Cazier,~N.; Mennemanteuil,~M.-M.;
  Berthelot,~J.; Smetanin,~I.~V.; Protsenko,~I.~E.; {Colas-des-Francs},~G.;
  Bouhelier,~A. Spontaneous Hot-Electron Light Emission from Electron-Fed
  Optical Antennas. \emph{Nano Lett.} \textbf{2015}, \emph{15},
  5811--5818\relax
\mciteBstWouldAddEndPuncttrue
\mciteSetBstMidEndSepPunct{\mcitedefaultmidpunct}
{\mcitedefaultendpunct}{\mcitedefaultseppunct}\relax
\EndOfBibitem
\bibitem[Gamaly(2011)]{Gamaly11}
Gamaly,~E. \emph{Femtosecond Laser-Matter Interaction: Theory, Experiments and
  Applications}; Pan Stanford Publishing Pte. Ltd: Singapore, 2011\relax
\mciteBstWouldAddEndPuncttrue
\mciteSetBstMidEndSepPunct{\mcitedefaultmidpunct}
{\mcitedefaultendpunct}{\mcitedefaultseppunct}\relax
\EndOfBibitem
\bibitem[Kanavin \latin{et~al.}(1998)Kanavin, Smetanin, Isakov, Afanasiev,
  Chichkov, Wellegehausen, Nolte, Momma, and T\"unnermann]{kanavin98}
Kanavin,~A.~P.; Smetanin,~I.~V.; Isakov,~V.~A.; Afanasiev,~Y.~V.;
  Chichkov,~B.~N.; Wellegehausen,~B.; Nolte,~S.; Momma,~C.; T\"unnermann,~A.
  Heat transport in metals irradiated by ultrashort laser pulses. \emph{Phys.
  Rev. B} \textbf{1998}, \emph{57}, 14698--14703\relax
\mciteBstWouldAddEndPuncttrue
\mciteSetBstMidEndSepPunct{\mcitedefaultmidpunct}
{\mcitedefaultendpunct}{\mcitedefaultseppunct}\relax
\EndOfBibitem
\bibitem[Agranat \latin{et~al.}(2015)Agranat, Ashitkov, Ovchinnikov, Sitnikov,
  Yurkevich, Chefonov, man, Anisimov, and Fortov]{Agranat15}
Agranat,~M.~B.; Ashitkov,~S.~I.; Ovchinnikov,~A.~V.; Sitnikov,~D.~S.;
  Yurkevich,~A.~A.; Chefonov,~O.~V.; man,~L. T.~P.; Anisimov,~S.~I.;
  Fortov,~V.~E. Thermal Emission of Hot Electrons in a Metal. \emph{J. Exp.
  Theor. Phys. Lett.} \textbf{2015}, \emph{9}, 598--602\relax
\mciteBstWouldAddEndPuncttrue
\mciteSetBstMidEndSepPunct{\mcitedefaultmidpunct}
{\mcitedefaultendpunct}{\mcitedefaultseppunct}\relax
\EndOfBibitem
\bibitem[McNeill \latin{et~al.}(2001)McNeill, O’Connora, Adams, Barbara, and
  K{\"a}mmer]{Barbara01}
McNeill,~J.~D.; O’Connora,~D.~B.; Adams,~D.~M.; Barbara,~P.~F.;
  K{\"a}mmer,~S.~B. Field-Induced Photoluminescence Modulation of MEH-PPV under
  Near-Field Optical Excitation. \emph{J. Phys. Chem. B} \textbf{2001},
  \emph{105}, 76--82\relax
\mciteBstWouldAddEndPuncttrue
\mciteSetBstMidEndSepPunct{\mcitedefaultmidpunct}
{\mcitedefaultendpunct}{\mcitedefaultseppunct}\relax
\EndOfBibitem
\bibitem[Ashcroft and Mermin(1968)Ashcroft, and Mermin]{Ashcroft-Mermin1968}
Ashcroft,~N.~W.; Mermin,~N.~D. \emph{Solid State Physics}, 3rd ed.; Saunders
  College: Philadelphia, 1968\relax
\mciteBstWouldAddEndPuncttrue
\mciteSetBstMidEndSepPunct{\mcitedefaultmidpunct}
{\mcitedefaultendpunct}{\mcitedefaultseppunct}\relax
\EndOfBibitem
\bibitem[Roloff \latin{et~al.}(2017)Roloff, Klemm, Gronwald, Huber, Lupton, and
  Bange]{Roloff17}
Roloff,~L.; Klemm,~P.; Gronwald,~I.; Huber,~R.; Lupton,~J.~M.; Bange,~S. Light
  Emission from Gold Nanoparticles under Ultrafast Near-Infrared Excitation:
  Thermal Radiation, Inelastic Light Scattering, or Multiphoton Luminescence?
  \emph{Nano Lett.} \textbf{2017}, \emph{17}, 7914--7919, PMID: 29182344\relax
\mciteBstWouldAddEndPuncttrue
\mciteSetBstMidEndSepPunct{\mcitedefaultmidpunct}
{\mcitedefaultendpunct}{\mcitedefaultseppunct}\relax
\EndOfBibitem
\bibitem[Lombardo \latin{et~al.}(2005)Lombardo, Stathis, Linder, Pey, Palumbo,
  and Tung]{lombardo05}
Lombardo,~S.; Stathis,~J.~H.; Linder,~B.~P.; Pey,~K.~L.; Palumbo,~F.;
  Tung,~C.~H. Dielectric breakdown mechanisms in gate oxides. \emph{J. App.
  Phys.} \textbf{2005}, \emph{98}, 121301\relax
\mciteBstWouldAddEndPuncttrue
\mciteSetBstMidEndSepPunct{\mcitedefaultmidpunct}
{\mcitedefaultendpunct}{\mcitedefaultseppunct}\relax
\EndOfBibitem
\bibitem[Theis \latin{et~al.}(1984)Theis, DiMaria, Kirtley, and Dong]{theis84}
Theis,~T.~N.; DiMaria,~D.~J.; Kirtley,~J.~R.; Dong,~D.~W. Strong Electric Field
  Heating of Conduction-Band Electrons in Si02. \emph{Phys. Rev. Lett.}
  \textbf{1984}, \emph{52}, 1445--1448\relax
\mciteBstWouldAddEndPuncttrue
\mciteSetBstMidEndSepPunct{\mcitedefaultmidpunct}
{\mcitedefaultendpunct}{\mcitedefaultseppunct}\relax
\EndOfBibitem
\bibitem[diMaria and Stasiak(1989)diMaria, and Stasiak]{dimaria89}
diMaria,~D.~J.; Stasiak,~J.~W. Trap creation in silicon dioxide produced by hot
  electrons. \emph{J. App. Phys.} \textbf{1989}, \emph{65}, 2342--2356\relax
\mciteBstWouldAddEndPuncttrue
\mciteSetBstMidEndSepPunct{\mcitedefaultmidpunct}
{\mcitedefaultendpunct}{\mcitedefaultseppunct}\relax
\EndOfBibitem
\bibitem[DiMaria \latin{et~al.}(1993)DiMaria, Cartier, and Arnold]{Dimaria93}
DiMaria,~D.~J.; Cartier,~E.; Arnold,~D. Impact ionization, trap creation,
  degradation, and breakdown in silicon dioxide films on silicon. \emph{J. Appl
  . Phys.} \textbf{1993}, \emph{73}, 3367--3384\relax
\mciteBstWouldAddEndPuncttrue
\mciteSetBstMidEndSepPunct{\mcitedefaultmidpunct}
{\mcitedefaultendpunct}{\mcitedefaultseppunct}\relax
\EndOfBibitem
\bibitem[Groseneken \latin{et~al.}(1999)Groseneken, Dagraeve, Nigam, {Van den
  Bosh}, and Maes]{Maes99}
Groseneken,~G.; Dagraeve,~R.; Nigam,~T.; {Van den Bosh},~G.; Maes,~H.~E. Hot
  carrier degradation and time-dependent dielectric breakdown in oxides.
  \emph{Microelectr. Eng.} \textbf{1999}, \emph{49}, 27--40\relax
\mciteBstWouldAddEndPuncttrue
\mciteSetBstMidEndSepPunct{\mcitedefaultmidpunct}
{\mcitedefaultendpunct}{\mcitedefaultseppunct}\relax
\EndOfBibitem
\bibitem[Scott and Dumin(1995)Scott, and Dumin]{Dumin95}
Scott,~R.~S.; Dumin,~D.~J. The Transient Nature of Excess Low-Level Leakage
  Currents in Thin Oxides. \emph{J. Electrochem. Soc} \textbf{1995},
  \emph{142}, 586--590\relax
\mciteBstWouldAddEndPuncttrue
\mciteSetBstMidEndSepPunct{\mcitedefaultmidpunct}
{\mcitedefaultendpunct}{\mcitedefaultseppunct}\relax
\EndOfBibitem
\bibitem[Balk(1988)]{Balk}
Balk,~P. In \emph{The Si-SiO2 System}; Balk,~P., Ed.; Elsevier, 1988\relax
\mciteBstWouldAddEndPuncttrue
\mciteSetBstMidEndSepPunct{\mcitedefaultmidpunct}
{\mcitedefaultendpunct}{\mcitedefaultseppunct}\relax
\EndOfBibitem
\bibitem[Shamir \latin{et~al.}(1997)Shamir, Mihaychuk, and van
  Driel]{vanDriel97}
Shamir,~N.; Mihaychuk,~J.~G.; van Driel,~H.~M. Transient charging and slow
  trapping in ultrathin SiO 2 films on Si during electronbombardment. \emph{J.
  Vac. Sci. Technol. A} \textbf{1997}, \emph{15}, 2081--2084\relax
\mciteBstWouldAddEndPuncttrue
\mciteSetBstMidEndSepPunct{\mcitedefaultmidpunct}
{\mcitedefaultendpunct}{\mcitedefaultseppunct}\relax
\EndOfBibitem
\bibitem[Kang \latin{et~al.}(1999)Kang, Buh, Lee, Kim, Im, and Kuk]{Kang99}
Kang,~C.~J.; Buh,~G.~H.; Lee,~S.; Kim,~C.~K.; Im,~K. M. M.~C.; Kuk,~Y. Charge
  trap dynamics in a SiO 2 layer on Si by scanning capacitance microscopy.
  \emph{Appl. Phys. Lett.} \textbf{1999}, \emph{74}, 1815--1817\relax
\mciteBstWouldAddEndPuncttrue
\mciteSetBstMidEndSepPunct{\mcitedefaultmidpunct}
{\mcitedefaultendpunct}{\mcitedefaultseppunct}\relax
\EndOfBibitem
\bibitem[Kumar \latin{et~al.}(2018)Kumar, Viarbitskaya, Cuche, Girard,
  Bolisetty, Mezzenga, Colas~des Francs, Bouhelier, and Dujardin]{Kumar18}
Kumar,~U.; Viarbitskaya,~S.; Cuche,~A.; Girard,~C.; Bolisetty,~S.;
  Mezzenga,~R.; Colas~des Francs,~G.; Bouhelier,~A.; Dujardin,~E. Designing
  Plasmonic Eigenstates for Optical Signal Transmission in Planar Channel
  Devices. \emph{ACS Phot.} \textbf{2018}, \emph{5}, 2328--2335\relax
\mciteBstWouldAddEndPuncttrue
\mciteSetBstMidEndSepPunct{\mcitedefaultmidpunct}
{\mcitedefaultendpunct}{\mcitedefaultseppunct}\relax
\EndOfBibitem
\bibitem[Harutyunyan \latin{et~al.}(2015)Harutyunyan, Martinson, Rosenmann,
  Khorashad, Besteiro, Govorov, and Wiederrecht]{Wiederrecht15}
Harutyunyan,~H.; Martinson,~A. B.~F.; Rosenmann,~D.; Khorashad,~L.~K.;
  Besteiro,~L.~V.; Govorov,~A.~O.; Wiederrecht,~G.~P. Anomalous ultrafast
  dynamics of hot plasmonicelectrons in nanostructures with hot spots.
  \emph{Nature Nanotech.} \textbf{2015}, \emph{10}, 770--774\relax
\mciteBstWouldAddEndPuncttrue
\mciteSetBstMidEndSepPunct{\mcitedefaultmidpunct}
{\mcitedefaultendpunct}{\mcitedefaultseppunct}\relax
\EndOfBibitem
\bibitem[Nicholls \latin{et~al.}(2019)Nicholls, Stefaniuk, Nasir,
  Rodr\'iguez-Fortu\~no, Wurtz, and Zayats]{Zayats19}
Nicholls,~L.~H.; Stefaniuk,~T.; Nasir,~M.~E.; Rodr\'iguez-Fortu\~no,~F.~J.;
  Wurtz,~G.~A.; Zayats,~A.~V. Designer photonic dynamics by using non-uniform
  electron temperature distribution for on-demand all-optical switching times.
  \emph{Nature Comm.} \textbf{2019}, \emph{10}, 2967\relax
\mciteBstWouldAddEndPuncttrue
\mciteSetBstMidEndSepPunct{\mcitedefaultmidpunct}
{\mcitedefaultendpunct}{\mcitedefaultseppunct}\relax
\EndOfBibitem
\end{mcitethebibliography}

\renewcommand{\bibsection}{}
\newcommand{\enquote}[1]{``#1''}
\footnotesize\singlespace
\providecommand{\latin}[1]{#1}
\makeatletter
\providecommand{\doi}
  {\begingroup\let\do\@makeother\dospecials
  \catcode`\{=1 \catcode`\}=2 \doi@aux}
\providecommand{\doi@aux}[1]{\endgroup\texttt{#1}}
\makeatother
\providecommand*\mcitethebibliography{\thebibliography}
\csname @ifundefined\endcsname{endmcitethebibliography}
  {\let\endmcitethebibliography\endthebibliography}{}


\end{document}